\documentclass[letterpaper, 10 pt, conference]{ieeeconf}  % Comment this line out if you need a4paper

\IEEEoverridecommandlockouts                              % This command is only needed if 
\usepackage{graphics} % for pdf, bitmapped graphics files
\usepackage{epsfig} % for postscript graphics files
\usepackage{mathptmx} % assumes new font selection scheme installed
\usepackage{times} % assumes new font selection scheme installed
\usepackage{amsmath} % assumes amsmath package installed
\usepackage{amssymb}  % assumes amsmath package installed
\usepackage{algorithm}
\usepackage{algpseudocode}
\usepackage{listings}
\usepackage{subcaption}
\usepackage{booktabs}
\usepackage{pgf, tikz, framed, url}
\usepackage[utf8]{inputenc}
\usepackage{placeins}  % Add this in the preamble
\usepackage{dblfloatfix}
\usepackage{caption}

\definecolor{codegreen}{rgb}{0,0.6,0}
\definecolor{codegray}{rgb}{0.5,0.5,0.5}
\definecolor{codepurple}{rgb}{0.58,0,0.82}
\definecolor{backcolour}{rgb}{0.95,0.95,0.92}

\title{\LARGE \bf
YOLO-MARL: You Only LLM Once for Multi-Agent Reinforcement Learning
}

\author{Yuan Zhuang$^{1\dag}$, Yi Shen$^{2\dag}$, Zhili Zhang$^{1}$, Yuxiao Chen$^{3}$ and Fei Miao$^{1*}$% <-this % stops a space
\thanks{$^{\dag}$ These authors contributed equally to this work.}% <-this % stops a space
\thanks{$^{1}$University of Connecticut}%
\thanks{$^{2}$University of Pennsylvania}%
\thanks{$^{3}$NVIDIA}%
\thanks{$^{*}$Corresponding author: Fei Miao (\texttt{fei.miao@uconn.edu})}%
}

\usepackage{cite}    
\makeatletter
\let\NAT@parse\undefined
\makeatother
\usepackage{hyperref}

\begin{document}

\maketitle
\thispagestyle{empty}
\pagestyle{empty}

%%%%%%%%%%%%%%%%%%%%%%%%%%%%%%%%%%%%%%%%%%%%%%%%%%%%%%%%%%%%%%%%%%%%%%%%%%%%%%%%
\begin{abstract}

Advancements in deep multi-agent reinforcement learning (MARL) have positioned it as a promising approach for decision-making in cooperative games. However, it still remains challenging for MARL agents to learn cooperative strategies for some game environments. Recently, large language models (LLMs) have demonstrated emergent reasoning capabilities, making them promising candidates for enhancing coordination among the agents. However, due to the model size of LLMs, it can be expensive to frequently infer LLMs for actions that agents can take. In this work, we propose You Only LLM Once for MARL (YOLO-MARL), a novel framework that leverages the high-level task planning capabilities of LLMs to improve the policy learning process of multi-agents in cooperative games. Notably, for each game environment, YOLO-MARL only requires one time interaction with LLMs in the proposed strategy generation, state interpretation and planning function generation modules,  before the MARL policy training process. This avoids the ongoing costs and computational time associated with frequent LLMs API calls during training. Moreover, trained decentralized policies based on normal-sized neural networks operate independently of the LLM. We evaluate our method across two different environments and demonstrate that YOLO-MARL outperforms traditional MARL algorithms. The Github repository of our code can be found at \href{https://github.com/paulzyzy/YOLO-MARL}{\color{blue}https://github.com/paulzyzy/YOLO-MARL}.
\end{abstract}
%%%%%%%%%%%%%%%%%%%%%%%%%%%%%%%%%%%%%%%%%%%%%%%%%%%%%%%%%%%%%%%%%%%%%%%%%%%%
\section{INTRODUCTION}
MARL has become a vital framework for solving decision-making problems in complex multi-agent systems. With the rising applications of multi-agent systems, it is increasingly crucial for individual agents to cooperate or compete in dynamic environments without relying on centralized control\cite{Benchmarking}. In cooperative Markov games, agents work together to maximize joint rewards. However, existing MARL approaches often struggle with learning effective distributed policies, particularly in tasks characterized by sparse rewards, dynamic environments, and large action spaces.

Concurrently, LLMs have demonstrated remarkable capabilities as high-level semantic planners by leveraging in-context learning and extensive prior knowledge\cite{ahn2022can}. Recent studies have showcased LLMs deployed as embodied agents \cite{zhang2023building, smartllm}, as well as their use in guiding reinforcement learning (RL) as ELLM, which leverages LLMs to suggest goals \cite{du2023guiding}, and work focusing on aligning LLM-provided actions with RL policies \cite{Kwon2023RewardDesign}. Despite these promising developments, integrating LLMs into multi-agent settings remains largely unexplored. Moreover, repeated interactions with LLMs during long-episode or complex tasks can be both time-consuming and computationally prohibitive, particularly when training over tens of millions of steps.

\begin{figure*}[ht]
\begin{center}
\includegraphics[width=0.8\linewidth]{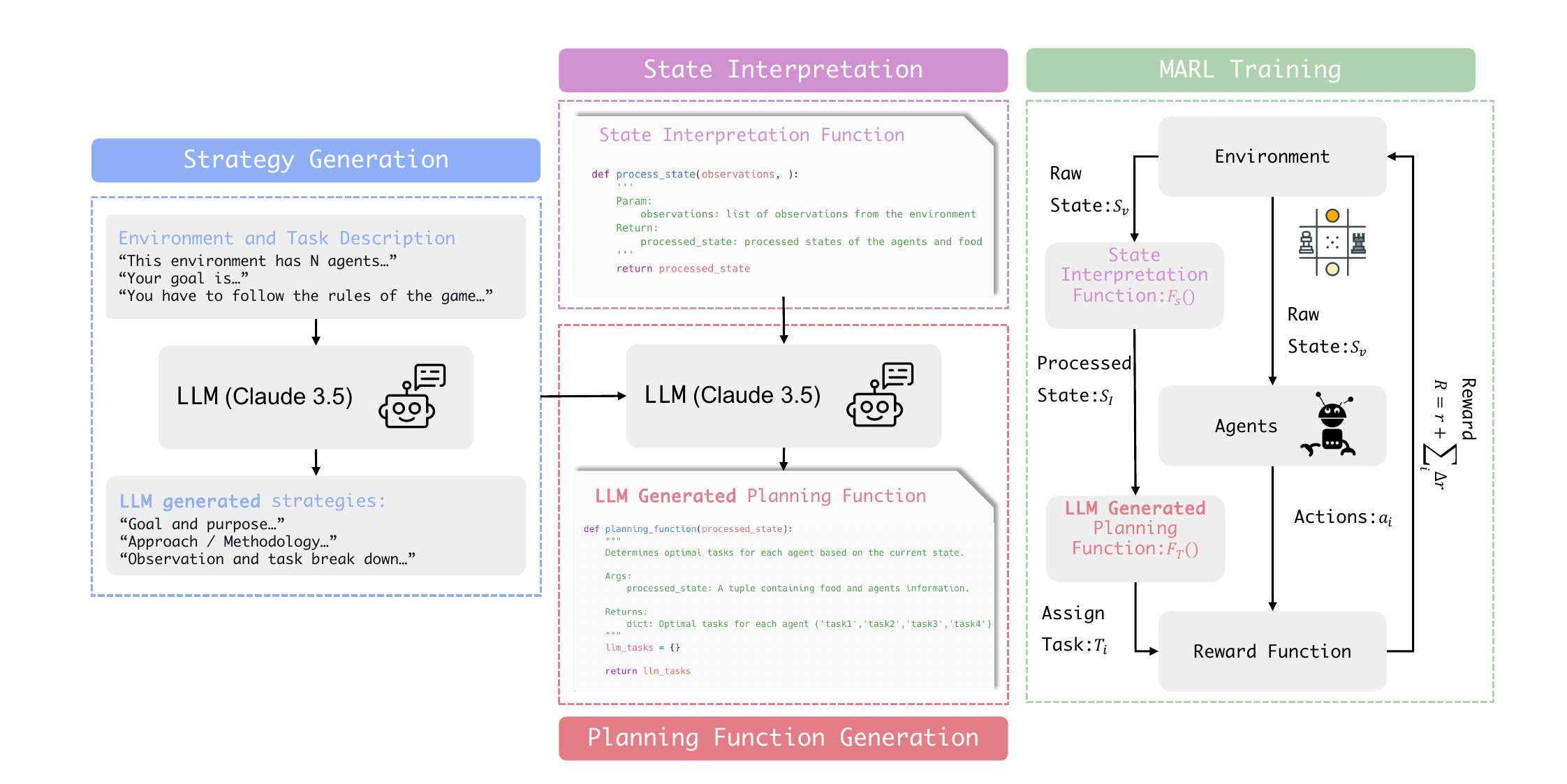}
\end{center}
\caption{Depiction of our framework YOLO-MARL. (a). Strategy Generation: We pass basic environment and task description into the LLM to get generated strategies for this specific environment. (b). State Interpretation: We process the global states so that the format of global states will be more structured and organized for better comprehension by the LLM. (c). Planning Function Generation: We chain together the environment and task description, LLM generated strategies and state interpretation function. These prompts are then fed into the LLM to generate a Python planning function for this environment. (d). MARL Training: The state interpretation function and the generated planning function are integrated into the MARL training process. The LLM is no longer required for further interaction after the Planning Function Generation. The more detailed explanation of MARL Training part can be found in Algorithm ~\ref{alg:yolo}}
\label{fig:framework}
\end{figure*}

Addressing these challenges, we propose YOLO-MARL, a novel framework combining LLM's high-level planning capabilities with MARL policy training, as shown in Fig.~\ref{fig:framework}. YOLO-MARL uniquely requires just one LLM interaction per environment. Once the strategy generation, state interpretation, and planning function generation modules produce the necessary guidance, the MARL training process proceeds without further LLM involvement. This design substantially reduces the communication overhead and computational cost typically associated with LLM inferences. Moreover, YOLO-MARL demonstrates its strong generalization capability and simplicity for application, since it only requires a basic understanding of the new environments from users. We validate our framework on a sparse reward multi-agent environment: Level-Based Foraging (LBF) environment\cite{Benchmarking} as well as the Multi-Agent Particle (MPE) environment\cite{MADDPG}, and our results show that YOLO-MARL outperforms several MARL baselines, such as\cite{MAPPO, QMIX, MADDPG}. To the best of our knowledge, this work represents one of the first trials to fuse the high-level reasoning and planning abilities of LLMs with MARL, pointing a new direction for scalable and efficient multi-agent coordination \cite{llmfuturesurvey}.

In summary, our proposed method YOLO-MARL has the following advantages:
\begin{itemize}
    \item This framework synergizes the planning capabilities of LLMs with MARL to enhance the policy learning in challenging cooperative game environments. In particular, our approach exploits the LLM's wide-ranging reasoning ability to generate high-level planning functions to facilitate agents in coordination. 
    \item YOLO-MARL requires minimal LLMs involvement, which significantly reduces computational overhead and mitigates communication connection instability concerns when invoking LLMs during the training process.     
    \item Our approach leverages zero-shot prompting and can be easily adapted to various game environments, with only basic prior knowledge required from users.
\end{itemize}
An overview of YOLO-MARL is presented in Fig.~\ref{fig:framework}, and all related prompts, environments, and generated planning functions are available in our GitHub repository.
\section{Related Work}
\label{gen_inst}
\subsection{Multi-Agent Reinforcement Learning}
MARL has attracted significant attention for its potential to address complex, decentralized problems. A popular paradigm is centralized training with decentralized execution. Methods like QMIX\cite{QMIX} and MADDPG\cite{MADDPG} employ centralized critics during training to coordinate agents, while allowing independent execution at test time. In cooperative environments, COMA\cite{COMA} and VDN\cite{VDN} enable agents to share rewards and maximize joint returns. Despite these advances, many existing MARL algorithms struggle in sparse-reward settings and have difficulty learning fully cooperative policies. Moreover, only limited work has explored the integration of LLMs with MARL\cite{llmfuturesurvey}, leaving open questions about leveraging LLMs for multi-agent decision-making. 
\subsection{Large Language Models for RL and Decision-Making}
Recent studies have incorporated LLMs into the RL training process to enhance performance. For instance, \cite{du2023guiding} improve agent exploration by aligning LLM-suggested goals with observed behaviors, while \cite{carta2023grounding} generate actions conditioned on language-based prompts during online RL. Similarly, \cite{Kwon2023RewardDesign} uses LLMs to provide scalar rewards that guide training. However, many of these approaches focus on single-agent settings or require extensive interactions with LLMs during training. And \cite{graphllm2024} highlights the limitations of LLMs in handling complex low-level tasks. Other works, such as \cite{fan2022minedojo}, studies a multi-task RL problem. SayCan \cite{ahn2022can} grounds LLMs via value functions of pretrained skills to execute abstract commands on robots. \cite{liang2023code} finds that code-writing LLMs can be re-purposed to write robot policy code. Additionally, studies like \cite{EUREKA} and \cite{Text2Reward} exploit LLMs’ prior knowledge and code-generation capabilities to produce reward functions. In contrast, our approach leverages LLMs to generate planning functions, thereby enhancing MARL without continuous LLM involvement.
\subsection{Large Language Models for Multi-Agent Systems}
LLM-based multi-agent systems have been employed in diverse applications requiring varied agent roles and collaborative decision-making\cite{guo2024llmmasurvey, llmfuturesurvey}. Camel\cite{li2024camel} and MetaGPT\cite{hong2023metagpt} deploy multiple LLM agents for tasks like brainstorming and software development. SMART-LLM\cite{kannan2023smart} decomposes multi-robot task planning into subgoals, and Co-NavGPT\cite{yu2023co} uses LLMs as global planners for cooperative navigation. Other research has focused on applying LLMs in strategic gaming environments\cite{llmgamesurvey2024, mindagent2024, llmwerewolf2024, li2023theory, ma2023large}. Unlike these approaches, which use LLMs directly as agents or decision-makers, our method harnesses the planning capabilities of LLMs to train compact, efficient MARL policies.

% LLM-based Multi-Agent (LLM-MA) systems focus on diverse agent profiles, interactions, and collective decision-making. While this allows agents to collaborate on complex tasks, it also increases computational overhead due to the communication between LLMs \cite{guo2024llmmasurvey}, \cite{llmfuturesurvey}. Camel\cite{li2024camel} and MetaGPT\cite{hong2023metagpt} employ multiple LLM agents to accomplish tasks like brainstorming and software development. SMART-LLM\cite{kannan2023smart} decompose multi-robot task plans into subgoals for LLM to enable efficient execution, while Co-NavGPT\cite{yu2023co} integrates LLMs as global planners for cooperative navigation. Numerous studies have also focused on leveraging the decision-making capabilities of LLMs in complex computer game environments.\cite{llmgamesurvey2024}. \cite{mindagent2024} proposed an interactive framework and a novel environment that leverage LLMs as dispatchers for multi-agent system gaming. \cite{llmwerewolf2024} fine-tuned LLMs based on gameplay outcomes, enabling them to adapt and improve their decision-making within the strategic game. \cite{li2023theory} explore the use of LLMs in cooperative games within a text-based environment, and \cite{ma2023large} explores LLMs in the StarCraft II environment. In contrast, our method leverages the planning abilities of LLM to train better small-size neural network-based MARL policies instead of using LLMs directly as agents.
\section{Problem Formulation}
\textbf{Markov game} is defined as a multi-agent decision-making problem when the interaction between multiple agents affect the state dynamics of the entire system and the reward of each agent under certain conditions \cite{littman1994markov}. In this work, we consider a Markov game, or a stochastic game \cite{owen1982game} defined as a tuple $G := (\mathcal{N}, S, A, \{r^i\}_{i \in \mathcal{N}}, p, \gamma)$, where $\mathcal{N}$ is a set of $N$ agents, $S= S^1 \times \cdots \times S^N$ is the joint state space, $A = A^1 \times \cdots \times A^N$ is the joint action space,  with $(S^i, A^i)$ as the state space and action space of agent $i$, respectively, $\gamma \in [0,1)$ is the discounting factor\cite{littman1994markov,owen1982game}. The state transition $p: S \times A \rightarrow \Delta(S)$ is controlled by the current state and joint action, where $\Delta(S)$ represents the set of all probability distributions over the joint state space $S$. Each agent has a reward function, $r^i: S \times A \rightarrow \mathbb{R}$. At time $t$, agent $i$ chooses its action $a^i_t$ according to a policy $\pi^i: S \rightarrow \Delta(A^i)$. 

For each agent $i$, it attempts to maximize its expected sum of discounted rewards, i.e. its objective function $J^i(s,\pi) = \mathbb{E} \left[ \sum_{t=1}^{\infty} \gamma^{t-1} r_t^i(s_t,a_t) | s_1 = s, a_t \sim \pi(\cdot | {s}_t) \right]$. In the literature, MARL algorithms\cite{MADDPG,MAPPO,QMIX} have been designed to train neural network-based policies $\pi_i(\theta_i)$. For a cooperative game, one shared reward function for all agents is widely used during the training process, which is also considered in this work.
\section{Methodology}
\label{sec:4}
In this section, we introduce YOLO-MARL, a framework that leverages the high-level planning capabilities of LLMs to enhance MARL. YOLO-MARL integrates four key components: Strategy Generation, State Interpretation, Planning Function Generation, and MARL training process with the LLM generated Planning Function incorporated throughout.
\begin{algorithm}[H]
    \small
    \caption{YOLO-MARL Training Process}\label{alg:yolo}
    \begin{algorithmic}[1]
        \Require Large Language Model $LLM$, State Interpretation function $F_{\text{S}}$, MARL actor $\mathcal{A}$,  MARL algorithm $MARL_{alg}$, Initial Prompts $P_{init}$
        \State \textbf{Hyperparameters:} reward signal $r'$, penalty signal $p'$
        \State $P_{Strategy} \sim LLM(P_{init})$ \textcolor{codegreen}{ // Strategy Generation}
        \State $P = P_{init} + P_{Strategy} + F_{\text{S}}$  \textcolor{codegreen}{// Chaining all the prompt for Planning Function Generation}
        \State $\mathcal{F_{\mathcal{T}}}  \sim  LLM(P)$  \textcolor{codegreen}{// Planning Function Generation: Sample functions code from the LLM}
        \State \textbf{MARL training with generated planning function}
        \For{each training step}
            \State $S_{I} \gets F_{\text{S}}(S_{v})$ \textcolor{codegreen}{ // State Interpretation: Get processed global observation $S_{I}$ from $F_{\text{S}}$}
            \State $ \mathcal{T}_1, \mathcal{T}_2, \dotsc \gets \mathcal{F_{\mathcal{T}}}(S_{I})$ \textcolor{codegreen}{// Assign tasks $\mathcal{T}$ to each agent}
            \State $a_1, a_2, \dotsc \gets \mathcal{A}(S_{v})$ \textcolor{codegreen}{ // Output actions from the actor}
            \For{each agent $i$}
                \If{$a_i \in \mathcal{T}_i$}
                    \State $\Delta r_i \gets r'$
                \Else
                    \State $\Delta r_i \gets p'$
                \EndIf
            \EndFor
            \State $R \gets r + \sum_{i} \Delta r_i$ \textcolor{codegreen}{ // Compute final reward for critic: More details are in equation ~\ref{eq1},~\ref{eq2}}
            \State $\pi (\theta)= MARL_{alg}(R)$ \textcolor{codegreen}{ // Use $R$ as the final reward for MARL training}
        \EndFor
        \State \Return Trained MARL policy
\end{algorithmic}
\end{algorithm}
\subsection{Strategy Generation}
\label{sec:4.1}
To ensure applicability across diverse environments—even for users with limited domain knowledge—we incorporate a \emph{Strategy Generation Module}, as shown in the blue box of Figure~\ref{fig:framework}(a). In this module, the LLM receives basic environment details (e.g., task descriptions, rules, and constraints) along with a general guideline, and then autonomously outputs a detailed strategy in a prescribed format without requiring extensive human input or expertise. 

The Strategy Generation is crucial for several reasons:
\begin{itemize}
    \item Reducing User Burden: It alleviates the need for users to comprehensively understand new environments, saving time and effort. 
    \item Enhancing Generalization: It enables the framework to adapt to different environments with minimal prompt modifications. 
    \item Facilitating Planning Function Generation: The strategies serve as vital components in the prompts used for the Planning Function Generation Module. The results of using YOLO-MARL but without Strategy Generation Module are shown in ablation study ~\ref{sec:without_LLM_strategy}  .
\end{itemize}
The LLM-generated strategies are incorporated into the prompt alongside other necessary information to facilitate the subsequent planning function generation.
\subsection{State Interpretation}
\label{sec:4.2}
In many simulation environments, observations are provided as vectors, with each component encoded in a non-semantic format. Although such representations are effective for training deep RL models, they pose challenges for LLMs that require context to interpret each component correctly.

We propose the \emph{State Interpretation Module} to assist the LLM in interpreting the environment state. By providing a semantically meaningful representation of the state, the LLM can successfully generate executable planning functions for training. Formally, given the current environment state in vector form $S_{v}$, we define an interpretation function $F_{S}$ such that $F_{S}(S_{v}) \to S_{I}$, where $S_{I}$ provides more explicit and meaningful information about each state component. 

Recent works like \cite{EUREKA} and \cite{Text2Reward} have demonstrated the success of enhancing LLMs performance by providing relevant environment code. In the same manner, we include the interpretation function $F_{S}$ in the prompting pipeline, formatted as Python environment code as shown in the purple box in Figure~\ref{fig:framework}(b). The State Interpretation Module significantly reduces the risk of the LLM generating erroneous functions with outputs incompatible with the training procedures. An ablation study on the effectiveness of this module can be found in Sec~\ref{sec:ablation_state_func}.
\subsection{Planning Function Generation}
\label{sec:4.3}
A crucial component of our method is leveraging the LLM to perform high-level planning instead of handling low-level actions. We combine all the prompts from the previous modules and input them into the LLM. The LLM then generates a reasonable and executable planning function that can be directly utilized in the subsequent training process.

To be more concise, given any processed state $S_I$, we define an assignment planning function as $\mathcal{F_{\mathcal{T}}}(S_I) \to \mathcal{T}_i \in \mathcal{T}$, where $\mathcal{T}=\{\mathcal{T}_1,..., \mathcal{T}_n\}$ is a set of target assignments that each agent can take.  We define the assignment set $\mathcal{T}$ over the action space such that an action can belong to multiple assignments and vice versa. For example, if the assignment space is defined as \(\mathcal{T}=\{Landmark\_0, Landmark\_1\} \), and landmark 0 and landmark 1 are located at the top right and top left positions relative to the agent respectively, then taking the action ``UP" can be associated with both assignments. Conversely, we can have multiple actions correspond to an assignment. For instance, moving towards ``Landmark 0" may involve actions like ``UP" and ``RIGHT". 

The \emph{Planning Function Generation} will only be required once for each new environment you try to use. After you interact with the LLM to get generated planning function, you can directly use it in the later training process with different MARL algorithms. This is referred to the red module in Fig.~\ref{fig:framework}(c).
\subsection{MARL training with Planning function incorporation}
To incorporate the planning function into MARL training, we add an extra reward term to the original reward provided by environments. Specifically, we define the final reward $R$ used by the critic as: 
\begin{equation} \label{eq1}R = r + \sum_{i} \Delta r_i. \end{equation}

Here, \( r \) is the original reward from the environment. For each agent \( i \), \( \Delta r_i \) is an additional reward or penalty that determined based on whether the action taken by the agent aligns with the task assigned by the planning function. Specifically:
\begin{equation}\label{eq2} \Delta r_i = \begin{cases} r', \text{if $a_i \in \mathcal{T}_i$}\\
p', \text{if $a_i \notin \mathcal{T}_i$} \end{cases} \end{equation}

Notably, we don't need to interact with the LLM during the entire training process, nor do we need to call the planning function after the policy has been trained. The training process $MARL_{alg}(R)$ takes $R$ as the reward function, uses the same state and action space. We follow the standard MARL algorithms and evaluation metrics within the literature, such as \cite{MAPPO}, \cite{QMIX}, and \cite{MADDPG}. Our method,  as shown in the green box in Fig.~\ref{fig:framework}(d), is highly efficient compared to approaches that interact with LLMs throughout the whole training process or directly use LLMs as agents. In practice, using the LLM's API to generate the planning function incurs minimal cost—less than a dollar per environment—even when using the most advanced LLM APIs.
\section{Experiments}
\label{others}
In this section, we evaluate our method across two different environments: MPE and LBF. 
\subsection{Setup}
\textbf{Baselines.} In our experiments, we compare the MARL algorithm MADDPG\cite{MADDPG}, MAPPO\cite{MAPPO} and QMIX\cite{QMIX} and set default hyper-parameters according to the well-tuned performance of human-written reward, and fix that in all experiments on this task to do MARL training. Experiment hyper parameters are listed in Appendix.

\textbf{LLM} As one of the most advanced LLMs available at the time of experimentation, Claude 3.5 Sonnet exhibited enhanced reasoning and code generation abilities critical for our framework’s strategy formulation and planning functions. This performance advantage led us to select \texttt{claude-3-5-sonnet-20240620} as our primary LLM for all reported experiments.\footnote{\url{https://www.anthropic.com/news/claude-3-5-sonnet}}

\textbf{Metrics.} To assess the performance of our method, we use the mean return in evaluation for all other environments. During evaluation, we rely solely on the default return values provided by the environments for both the baseline and our method, ensuring a fair comparison.
\subsection{Results}
\label{sec:5.2}
\begin{figure*}[h]
    \centering
    \begin{subfigure}[b]{0.32\linewidth}
        \centering
        \includegraphics[width=\linewidth]{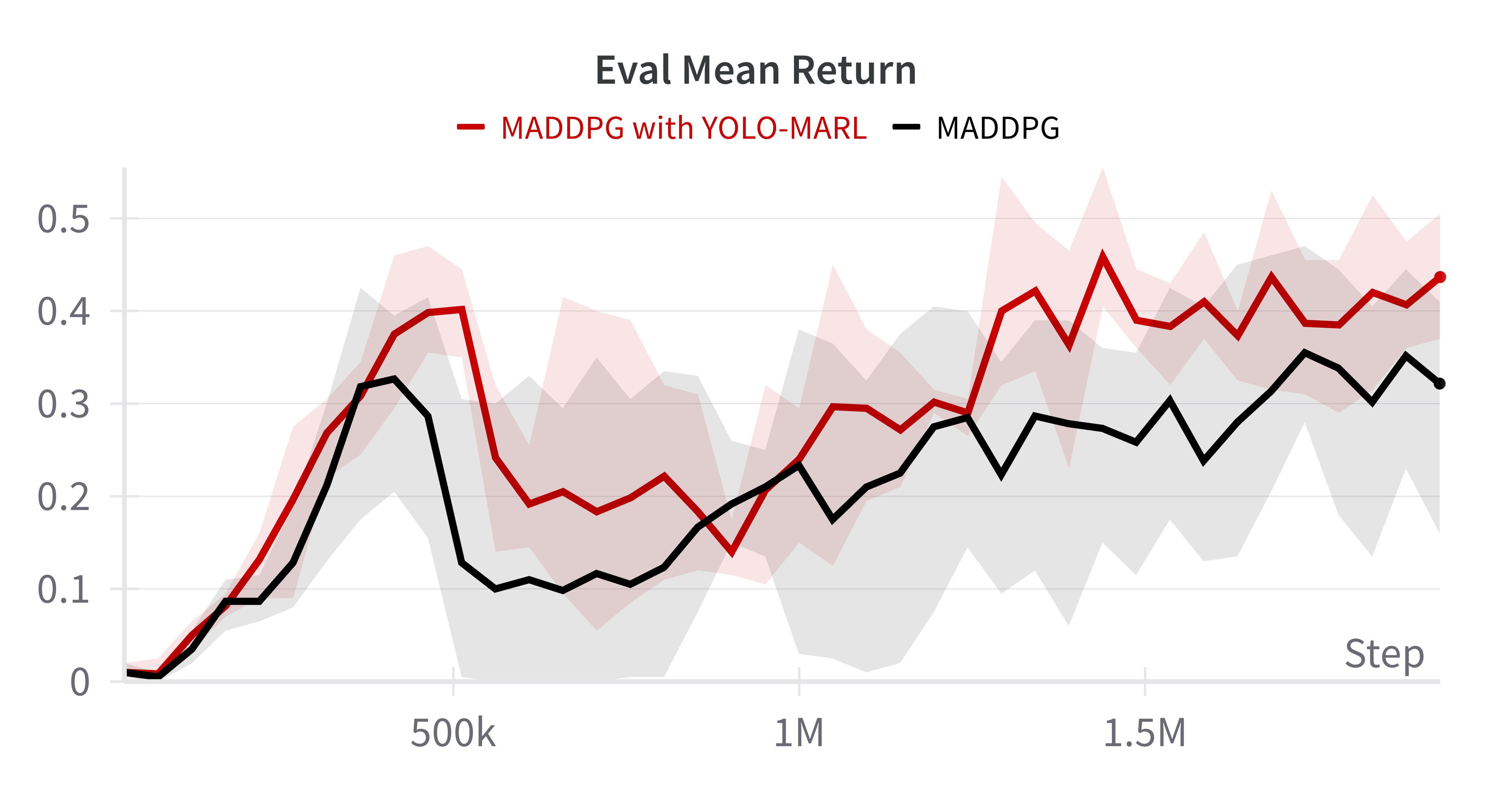}
        \caption{MADDPG}
        \label{fig:maddpg_lbf_3seeds}
    \end{subfigure}
    \hfill
    \begin{subfigure}[b]{0.32\linewidth}
        \centering
        \includegraphics[width=\linewidth]{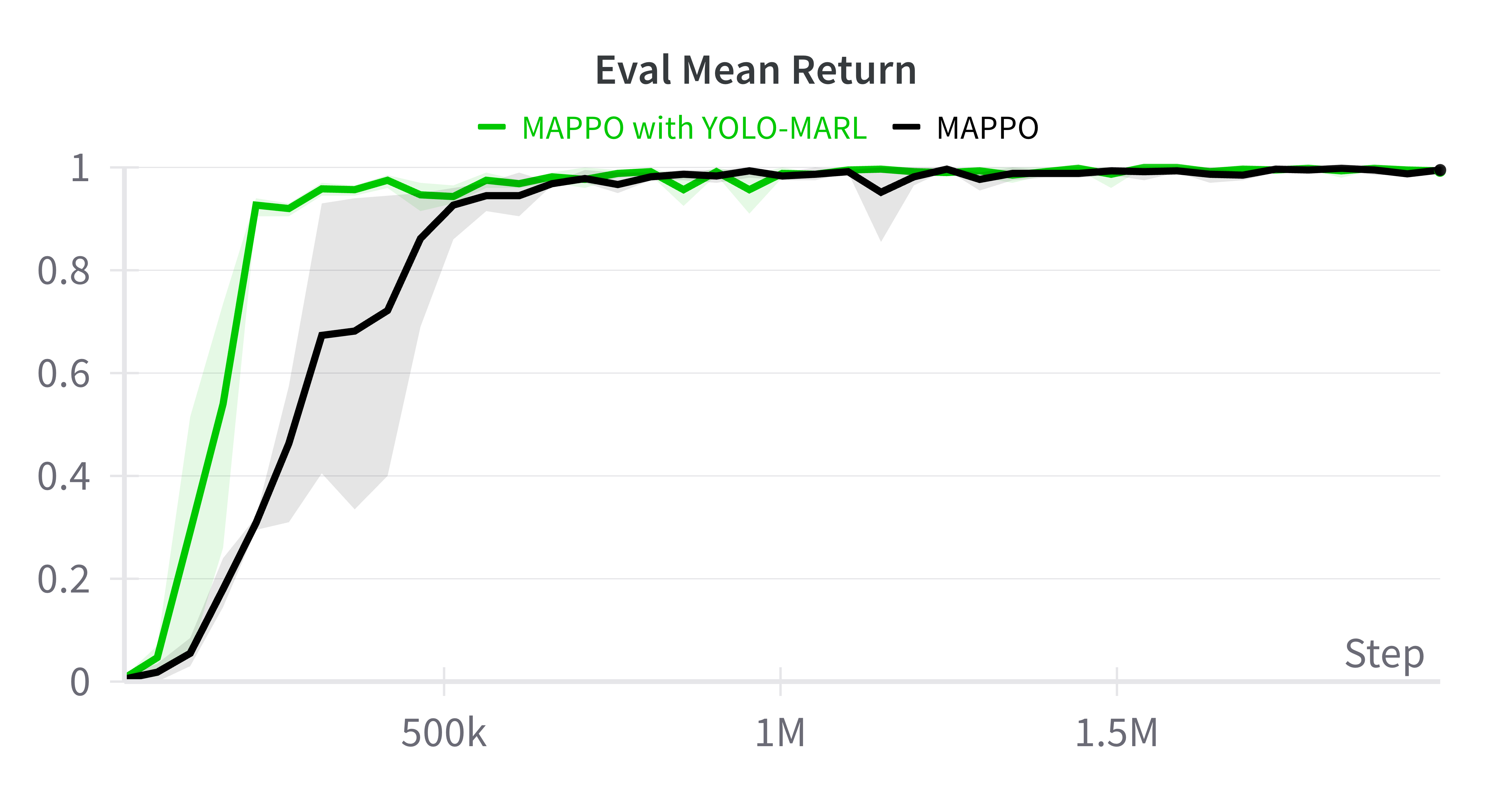}
        \caption{MAPPO}
        \label{fig:mappo_lbf_3seeds}
    \end{subfigure}
    \hfill
    \begin{subfigure}[b]{0.32\linewidth}
        \centering
        \includegraphics[width=\linewidth]{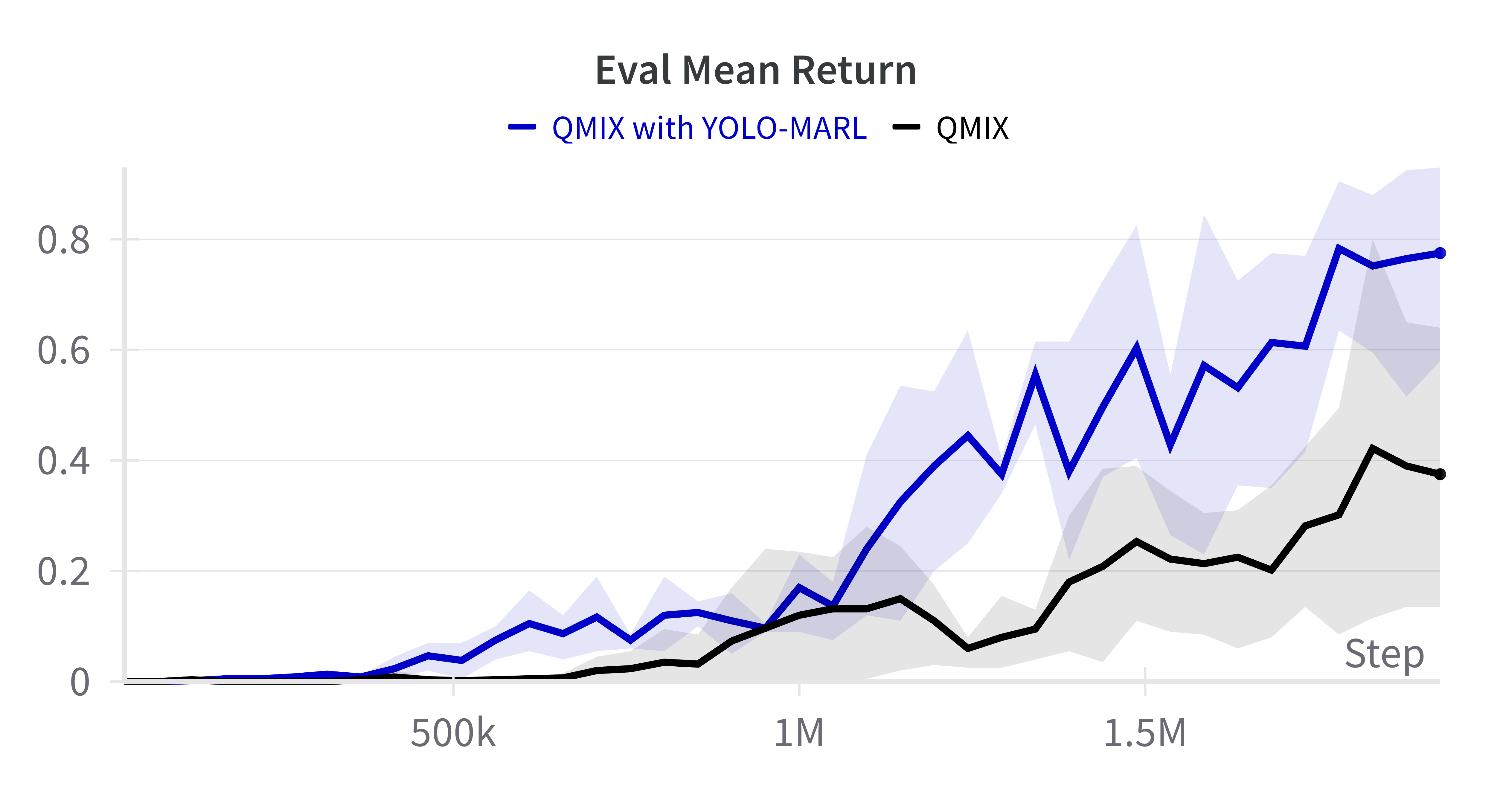}
        \caption{QMIX}
        \label{fig:qmix_lbf_3seeds}
    \end{subfigure}
    \caption{\textbf{Results for LBF environment across 3 seeds:}  The solid lines indicate the mean performance, and the shaded areas represent the range (minimum to maximum) across 3 different seeds.}
    \label{fig:comparison_LBF_3seeds}
\end{figure*}

\begin{table*}[t]
\caption{Comparison between YOLO-MARL and MARL in the LBF environment across three seeds. The highest evaluation return means during training are highlighted in bold. The corresponding results can be found in Figure~\ref{fig:comparison_LBF_3seeds}. The M means one million training steps. We run all the experiments on the same machine.}
\label{LBF_3seeds_table}
\begin{center}
\begin{tabular}{cccc}
 & \multicolumn{3}{c}{Mean Return after 0.2M / 0.4M / 1.5M / 2M Steps}\\
\cmidrule(lr){2-4}  % Proper ranges for \cmidrule
 & QMIX   & MADDPG & MAPPO \\
\midrule
MARL &  0.00/ 0.01/ 0.25/ 0.38 & 0.09/ 0.33/ 0.26/ 0.32 & 0.31/ 0.72/ 0.99/ 0.99\\
YOLO-MARL & \textbf{0.01/ 0.02 / 0.60/ 0.78} & \textbf{0.13/ 0.38/ 0.39/ 0.44} & \textbf{0.93/ 0.98/ 0.99/ 0.99}\\
\end{tabular}
\end{center}
\end{table*}
\textbf{Level-Based Foraging.} LBF\cite{Benchmarking} is a challenging sparse reward environment designed for MARL training. In this environment, agents must learn to navigate a path and successfully collect food, with rewards only being given upon task completion. To evaluate our framework in a cooperative setting, we selected the 2-player, 2-food fully cooperative scenario. In this setting, all agents must work together and coordinate their actions to collect the food simultaneously. The environment offers an action space consisting of [NONE, NORTH, SOUTH, WEST, EAST, LOAD], and we define the task set as [NONE, Food i, ..., LOAD]. Using the relative positions of agents and food items, we map assigned tasks to the corresponding actions in the action space and calculate the reward based on this alignment. We evaluated our framework over 3 different seeds, with the results shown in Figure~\ref{fig:comparison_LBF_3seeds} and Table~\ref{LBF_3seeds_table}. LLM assist the MARL algorithm by providing reward signals, our framework significantly outperformed the baseline, achieving a maximum improvement of \textbf{105 \%} in mean return and a \textbf{2x faster} convergence rate among all tested MARL algorithms. According to the results, our framework is effective across all the baseline algorithms, with particularly large improvements observed in QMIX and MADDPG, and a faster convergence rate for MAPPO. To assess the variability in the quality of our generated functions, we present the results of three different generated functions in Figure~\ref{fig:comparison_LBF_3F} and Table~\ref{LBF_3F_table} in Appendix. The results demonstrate that our framework consistently generates high-quality functions, with each achieving similar improvements across all baseline algorithms.

\begin{figure*}[h]
    \centering
    %----------- Row 1: 3 agents -----------
    \begin{subfigure}[b]{0.4\linewidth}
        \centering
        \includegraphics[width=\linewidth]{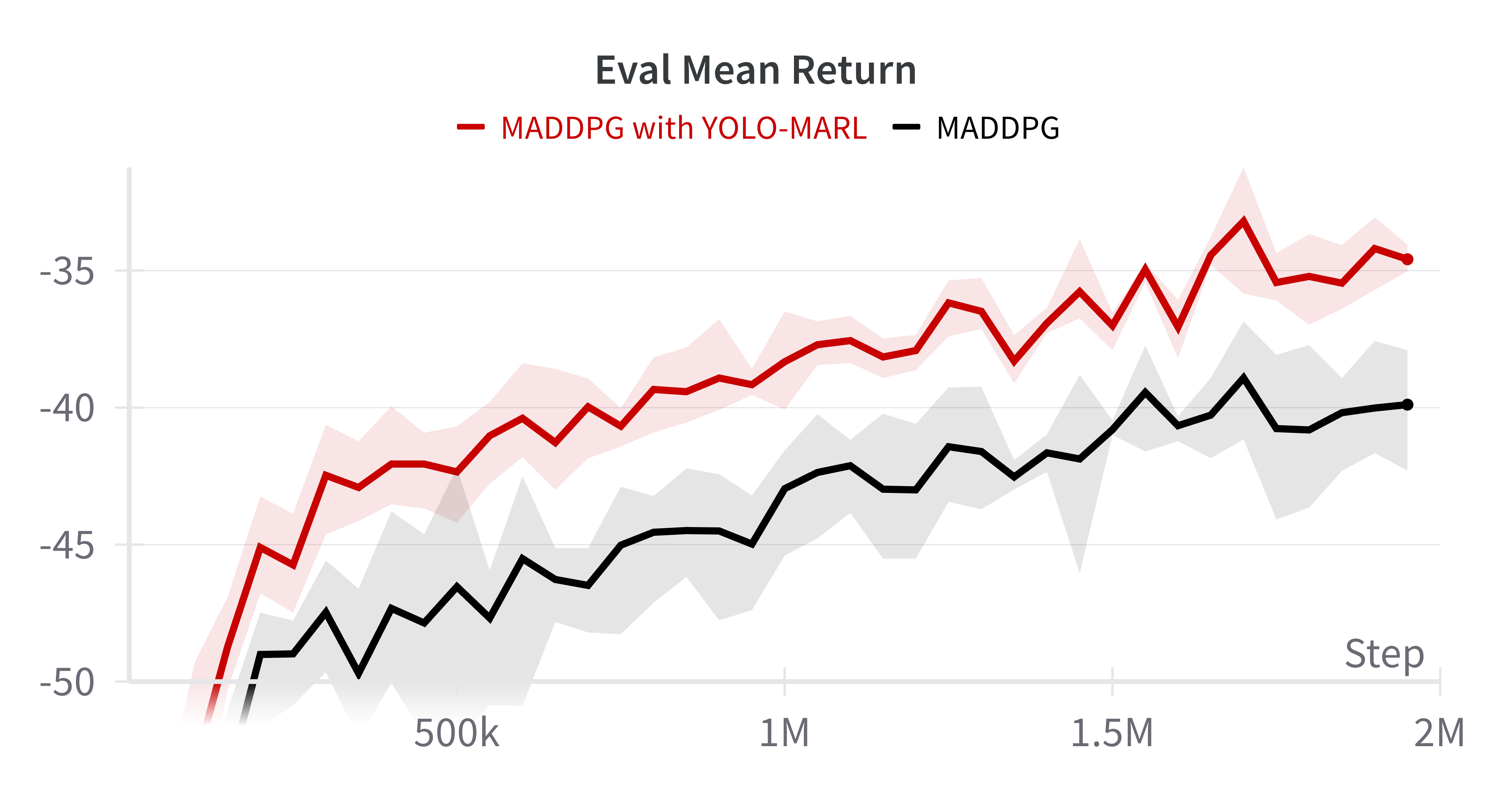}
        \caption{3 Agents -- MADDPG}
        \label{fig:maddpg_3_mpe}
    \end{subfigure}
    \hfill
    \begin{subfigure}[b]{0.4\linewidth}
        \centering
        \includegraphics[width=\linewidth]{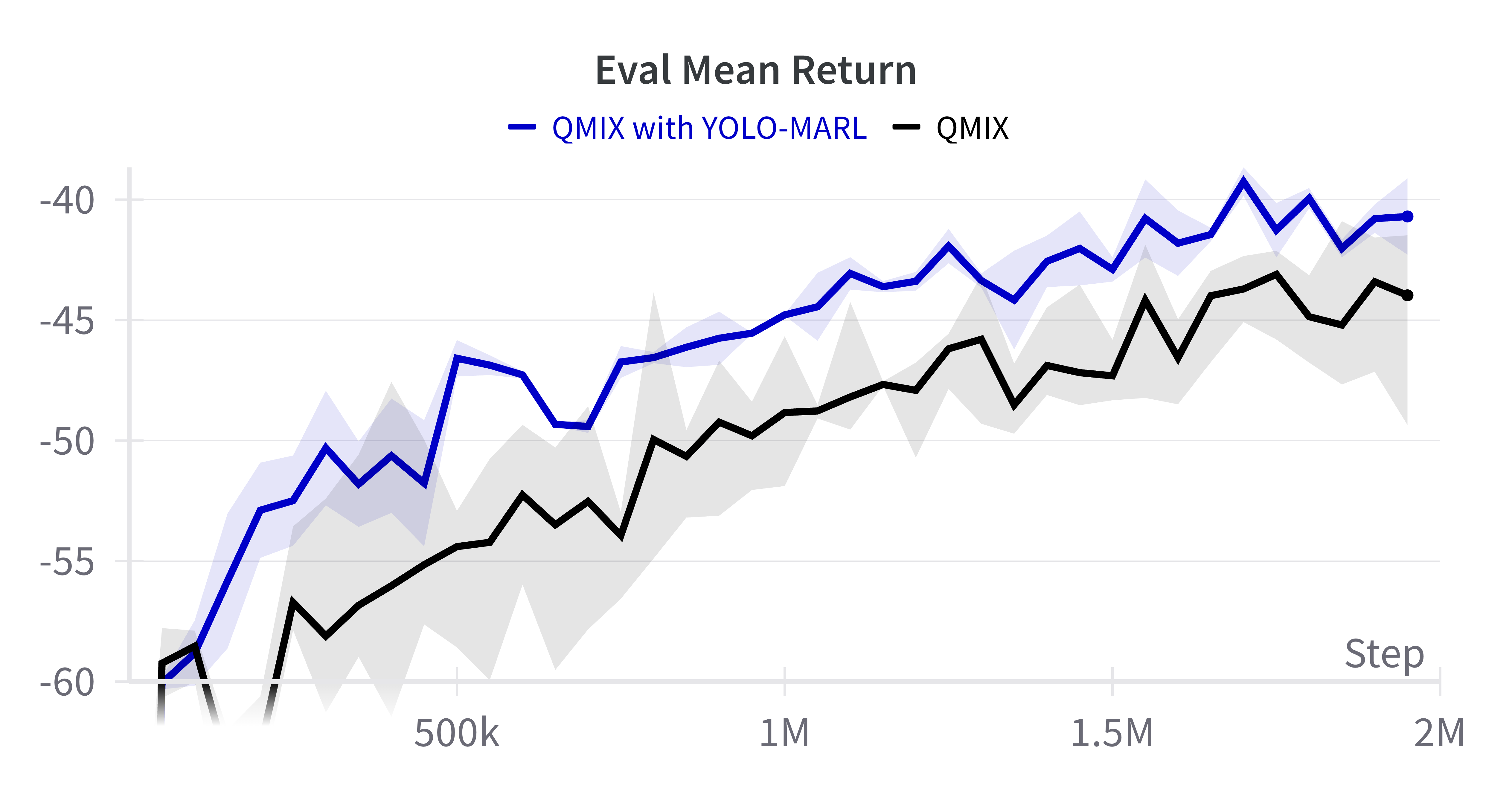}
        \caption{3 Agents -- QMIX}
        \label{fig:qmix_3_mpe}
    \end{subfigure}

    %----------- Row 2: 4 agents -----------
    \begin{subfigure}[b]{0.4\linewidth}
        \centering
        \includegraphics[width=\linewidth]{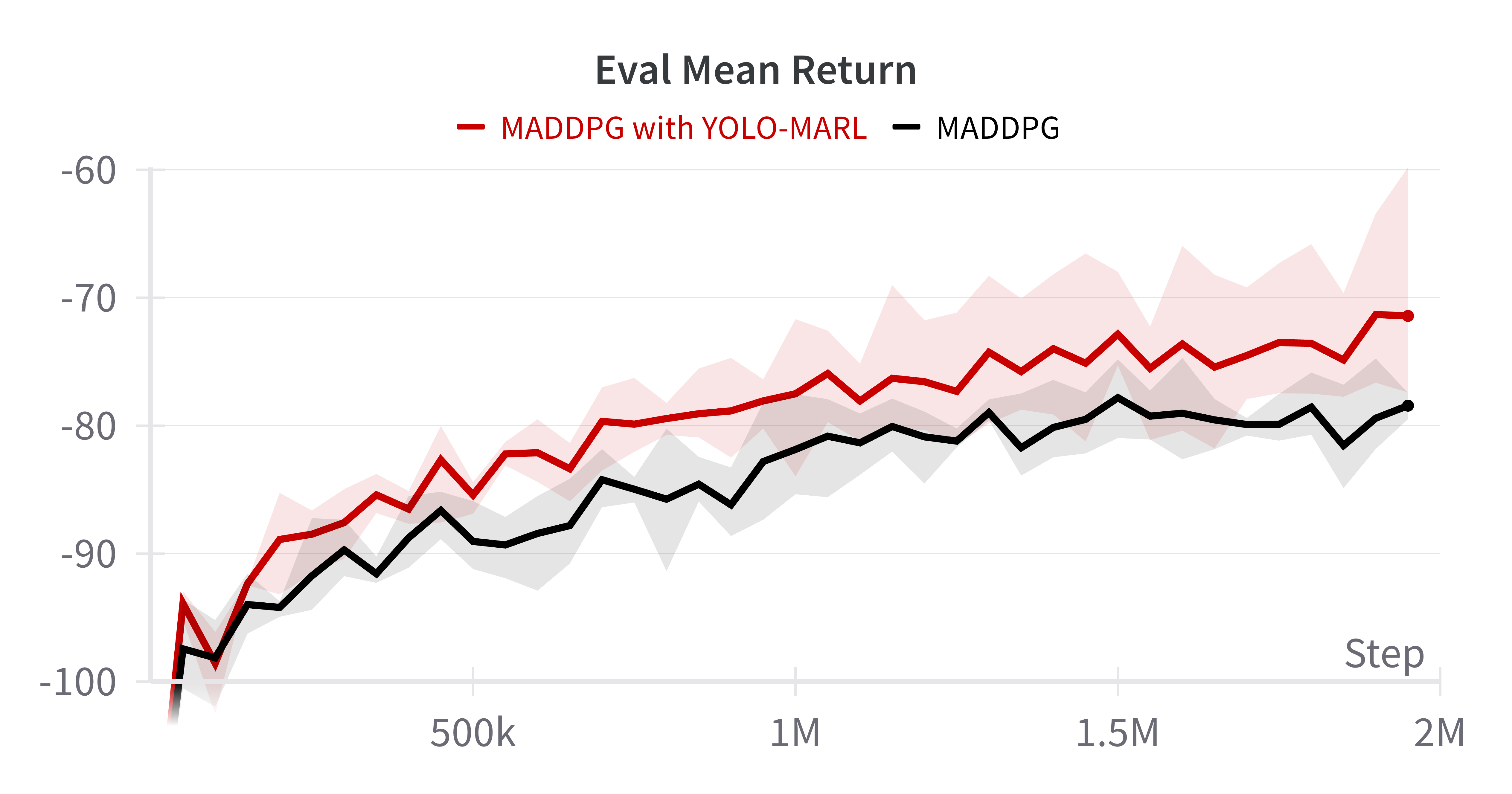}
        \caption{4 Agents -- MADDPG}
        \label{fig:maddpg_4_mpe}
    \end{subfigure}
    \hfill
    \begin{subfigure}[b]{0.4\linewidth}
        \centering
        \includegraphics[width=\linewidth]{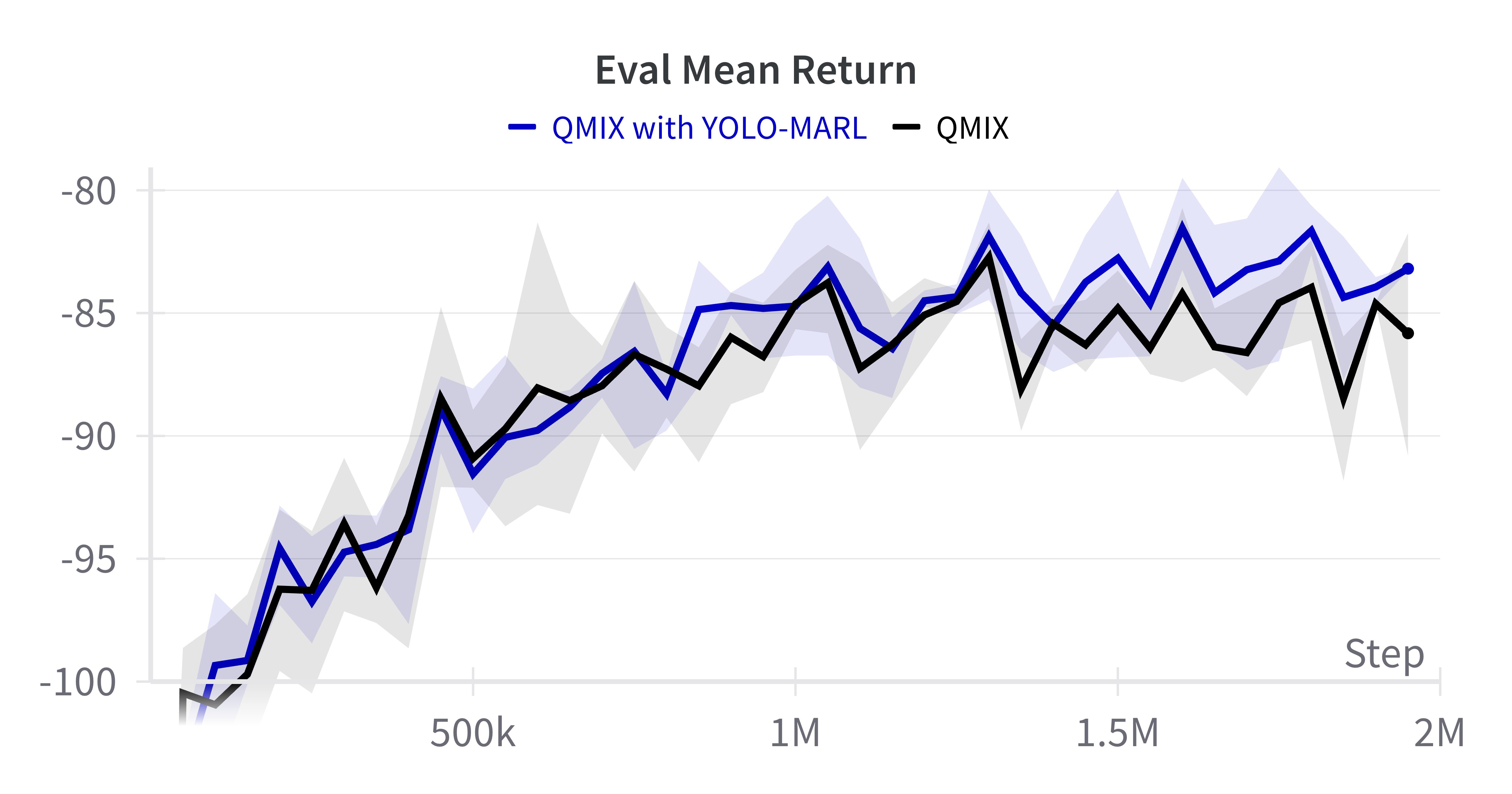}
        \caption{4 Agents -- QMIX}
        \label{fig:qmix_4_mpe}
    \end{subfigure}

    \caption{\textbf{Results for MPE simple spread environment} with 3 agents (top row a and b) and 4 agents (bottom row c and d).
    The solid lines indicate the mean performance, and the shaded areas represent the range (minimum to maximum) across 3 different generated planning functions.}
    \label{fig:comparison_MPE_3_4}
\end{figure*}

\label{sec:5.2.2}
\textbf{Multi-Agent Particle Environment.} We evaluate our framework in MPE\cite{MADDPG} simple spread environment which is a fully cooperative game. This environment has N agents, N landmarks. At a high level, agents must learn to cover all the landmarks while avoiding collisions. It's action space consists of [no\_action, move\_left, move\_right, move\_down, move\_up]. We define the assignment for each agent to take to be [Landmark\_i,...,No action]. During training, based on the global observation, we obtain the relative position of each agent with respect to the landmarks. Similar to LBF, we map each assignment of agent back to the corresponding action space and then reward the action of policy in action space level. We evaluate our approach on 3-agent and 4-agent scenarios using QMIX and MADDPG as baselines. As shown in Figure~\ref{fig:comparison_MPE_3_4}, our framework(colored line) outperforms the baseline(black line) algorithm in mean returns by \textbf{7.66\%}  and \textbf{8.8\%} for 3-agent scenario, and \textbf{2.4\%} and \textbf{18.09\%} for 4-agent scenario with QMIX and MADDPG respectively. These improvements demonstrate the effectiveness of our framework in enhancing coordination among agents to cover up all the landmarks.
\section{Ablation Study}
In this section, we conduct the ablation studies mainly in LBF 2 players 2 food fully cooperative environment since rewards in LBF are sparser compared to MPE\cite{Benchmarking}. We refer to ~\ref{sec:5.2} for more information about the environment.
\subsection{Comparison between YOLO-MARL with and without Strategy Generation}
\label{sec:without_LLM_strategy}
In this section, we examine the impact of the Strategy Generation Module on the performance of the YOLO-MARL framework. Specifically, we compare the standard YOLO-MARL with a variant that excludes the Strategy Generation Module to assess its significance.

\begin{figure*}[h]
    \centering
    \begin{subfigure}[b]{0.32\linewidth}
        \centering
        \includegraphics[width=\linewidth]{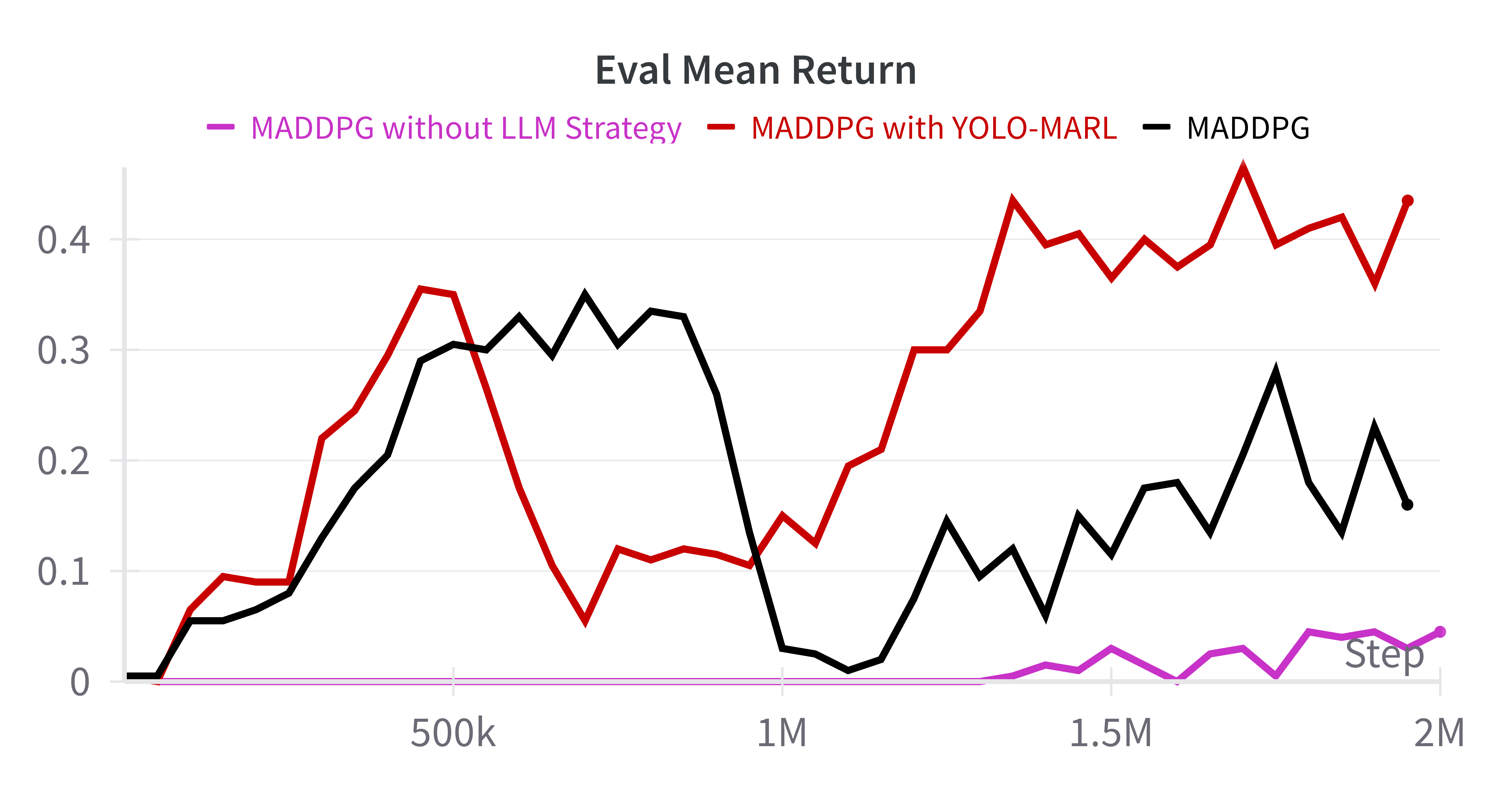}
        \caption{MADDPG}
        \label{fig:MADDPG_without_LLMstrategy}
    \end{subfigure}
    \hfill
    \begin{subfigure}[b]{0.32\linewidth}
        \centering
        \includegraphics[width=\linewidth]{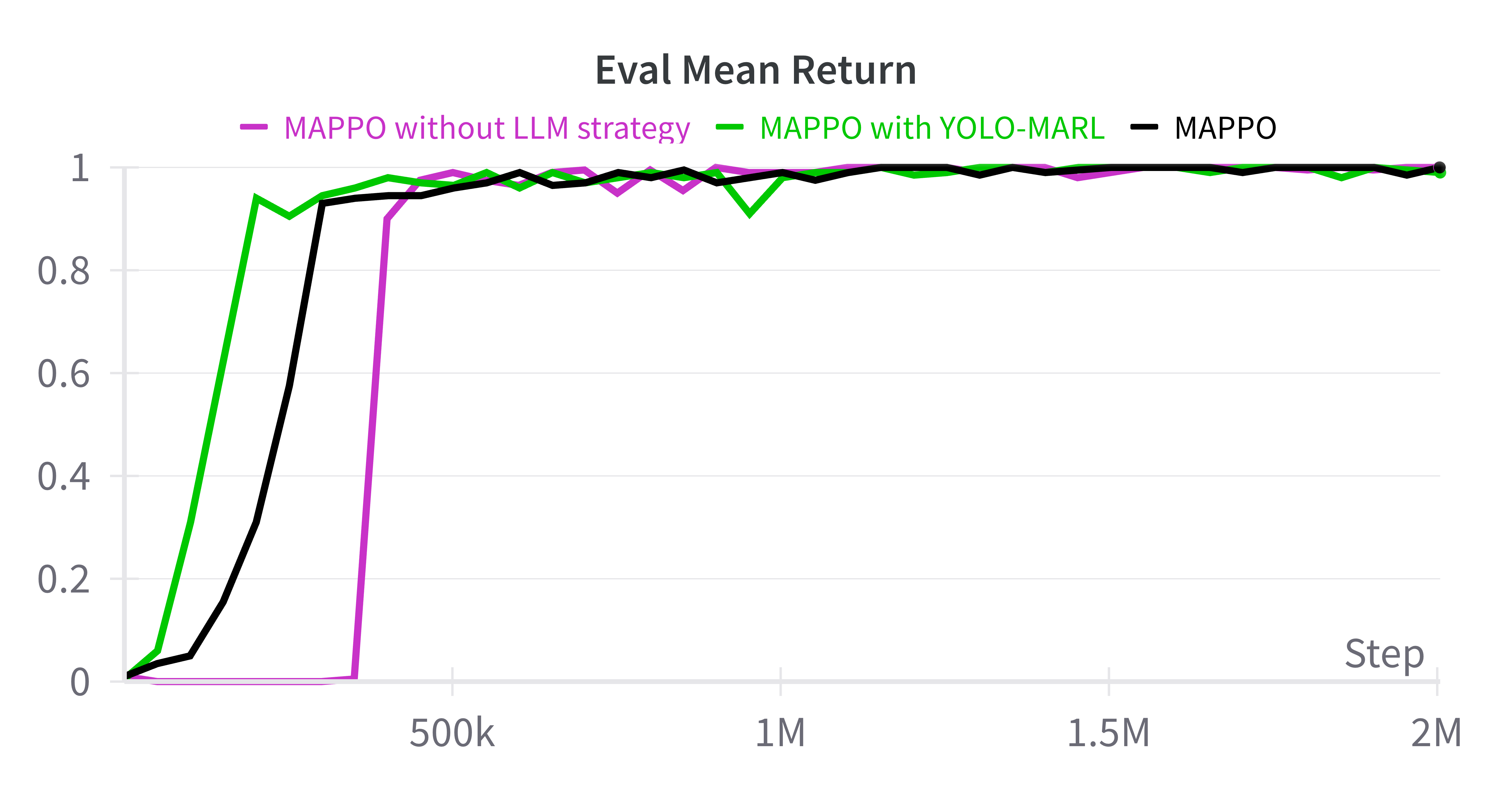}
        \caption{MAPPO}
        \label{fig:MAPPO_without_LLMstrategy}
    \end{subfigure}
    \hfill
    \begin{subfigure}[b]{0.32\linewidth}
        \centering
        \includegraphics[width=\linewidth]{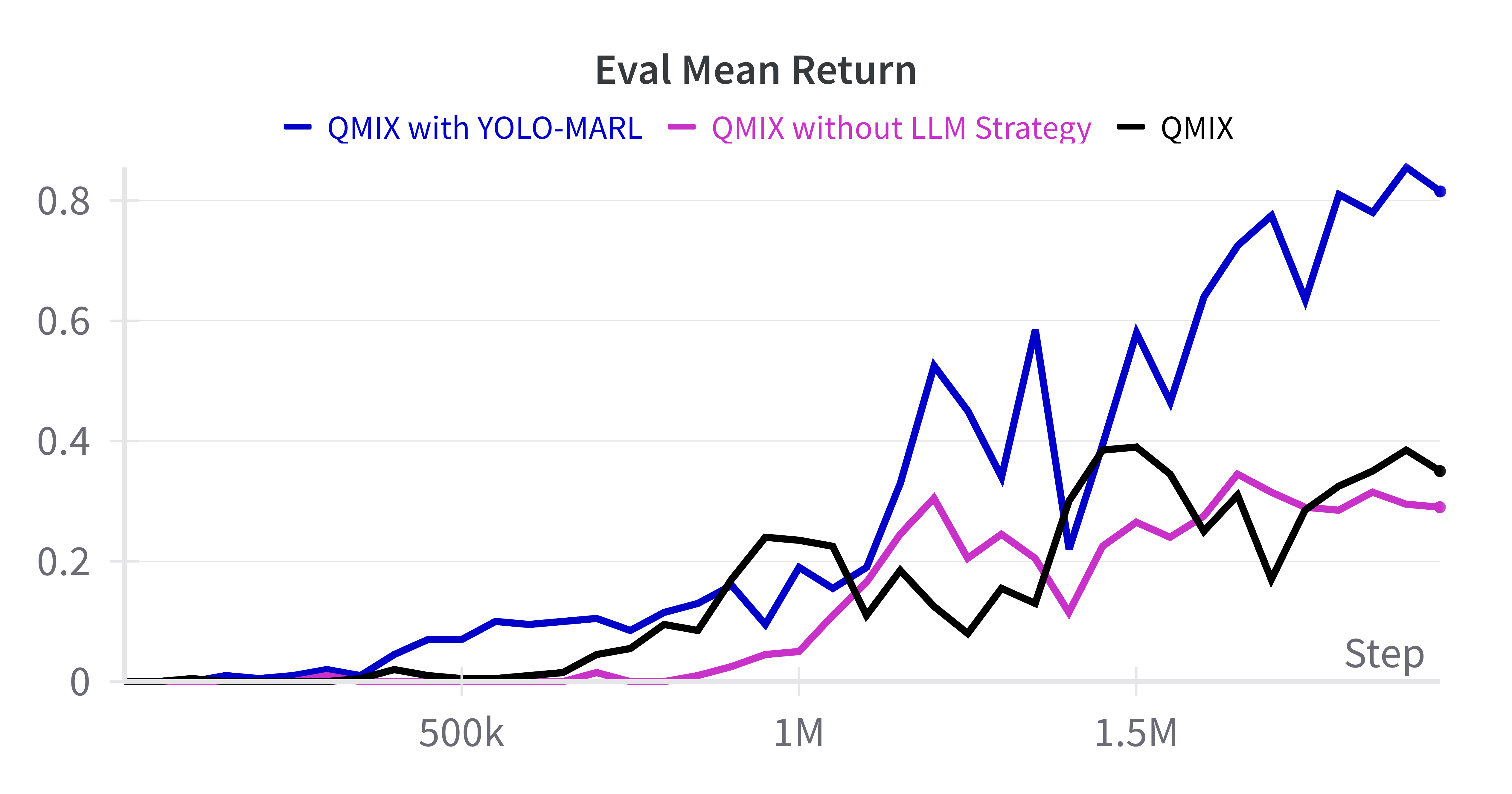}
        \caption{QMIX}
        \label{fig:QMIX_without_LLMstrategy}
    \end{subfigure}
    \caption{Comparison between YOLO-MARL with and without using LLM generated strategies in LBF}
    \label{fig:comparison_LBF_nostrategy}
\end{figure*}

According to our tests, the Strategy Generation Module plays an important role in the YOLO-MARL method. As shown in Figure~\ref{fig:comparison_LBF_nostrategy}
, without the LLM generated strategy, we obtain a worse-performing planning function. Interestingly, the mean returns of evaluations for the functions without the LLM generated strategy are not always close to zero, indicating that the generated planning functions are not entirely incorrect. Based on this, we could confirm that the Strategy Generation Module would help Planning Function Generation Module provides better solutions to this game. Moreover, giving the strategy also helps stabilize the quality of the generated code. We observe a higher risk of obtaining erroneous functions without supplying the strategy.
\subsection{Comparison between YOLO-MARL with and without State Interpretation}
\label{sec:ablation_state_func}
To demonstrate how the State Interpretation Module enhances our framework, we present two failure case snippets: 
\begin{itemize}
    \item Without the Interpretation Function: The interpretation function is omitted entirely from the prompting pipeline. 
    \item Providing Raw Environment Code Directly: Raw environment source code is fed directly to the LLM.
\end{itemize}

The LLM is unable to infer the type of state and attempts to fetch environment information via a non-existent key if no preprocessing code provided. And if environment code is provided without dimensional context for each component, the LLM is likely to make random guesses. In both scenarios, the absence of explicit state interpretation hinders the LLM's ability to generate accurate and executable planning functions. These failures underscore the importance of the State Interpretation Module in bridging the gap between vectorized observations and the LLM's requirement for semantically meaningful input.

By incorporating the State Interpretation Module, we enable the LLM to understand the environment's state representation effectively. This results in the generation of reliable planning functions that significantly enhance the performance of our YOLO-MARL framework.

\subsection{Comparison between YOLO-MARL and reward generation}
\label{sec:reward_generation_comparison}
In this section, we compare our YOLO-MARL method with approaches that utilize the LLM for reward generation without reward function template. We explore two scenarios: reward generation without feedback and reward generation with feedback. For the reward generation without feedback, the reward function is generated at the same stage as the planning function for fair comparison. This means that we generate the reward function before all the training process for each new environment. For the reward generation with feedback, we first generate a reward function just like the reward generation without feedback. And then, iteratively, we will run a whole training process on this environment and pass the feedback of this training performance to the LLM, combined with previous prompts and ask the LLM to refine the previous generated reward function.

\begin{figure*}[h]
    \centering
    \begin{subfigure}[b]{0.32\linewidth}
        \centering
        \includegraphics[width=\linewidth]{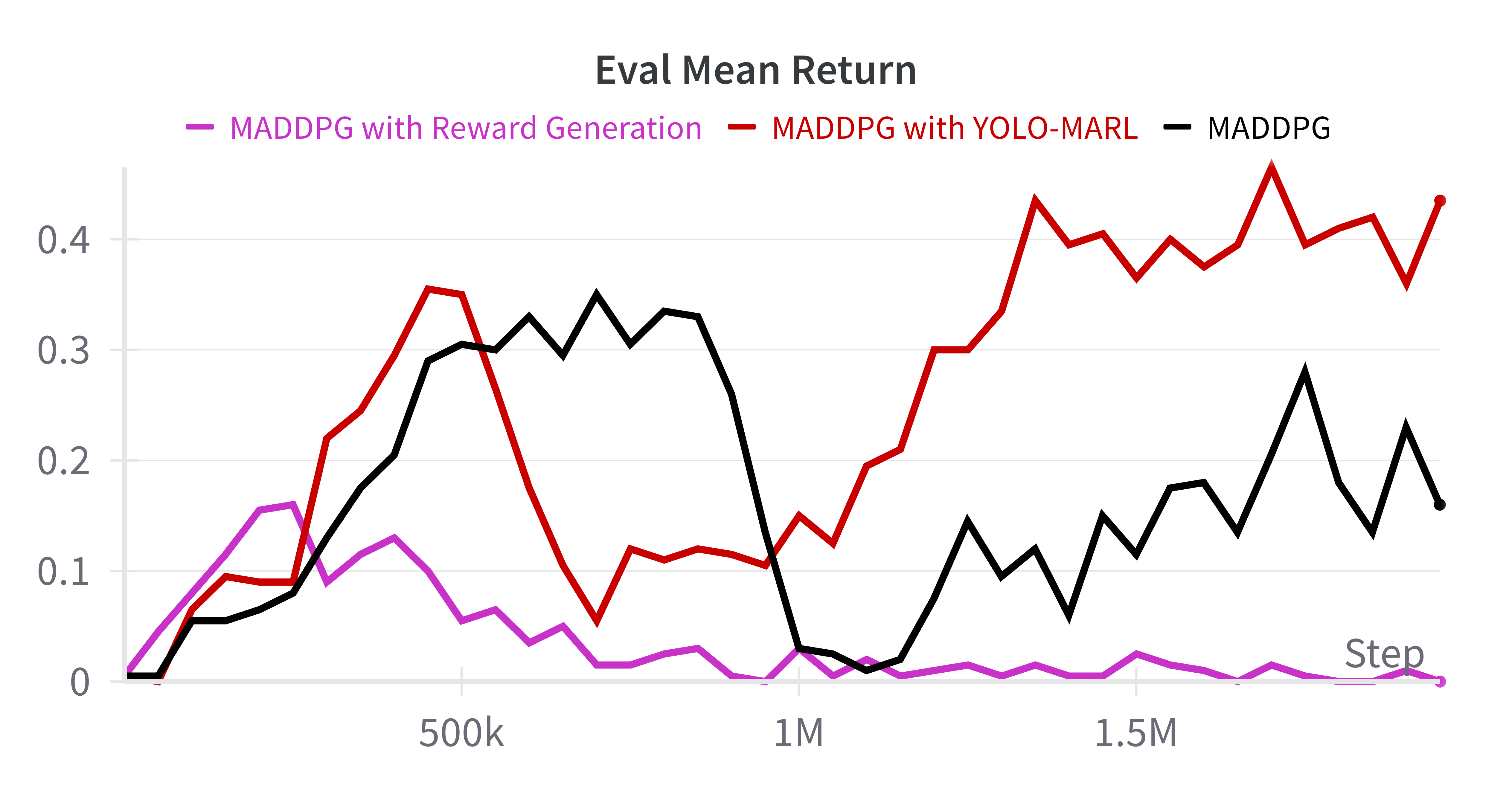}
        \caption{MADDPG}
        \label{fig:MADDPG_only_reward}
    \end{subfigure}
    \hfill
    \begin{subfigure}[b]{0.32\linewidth}
        \centering
        \includegraphics[width=\linewidth]{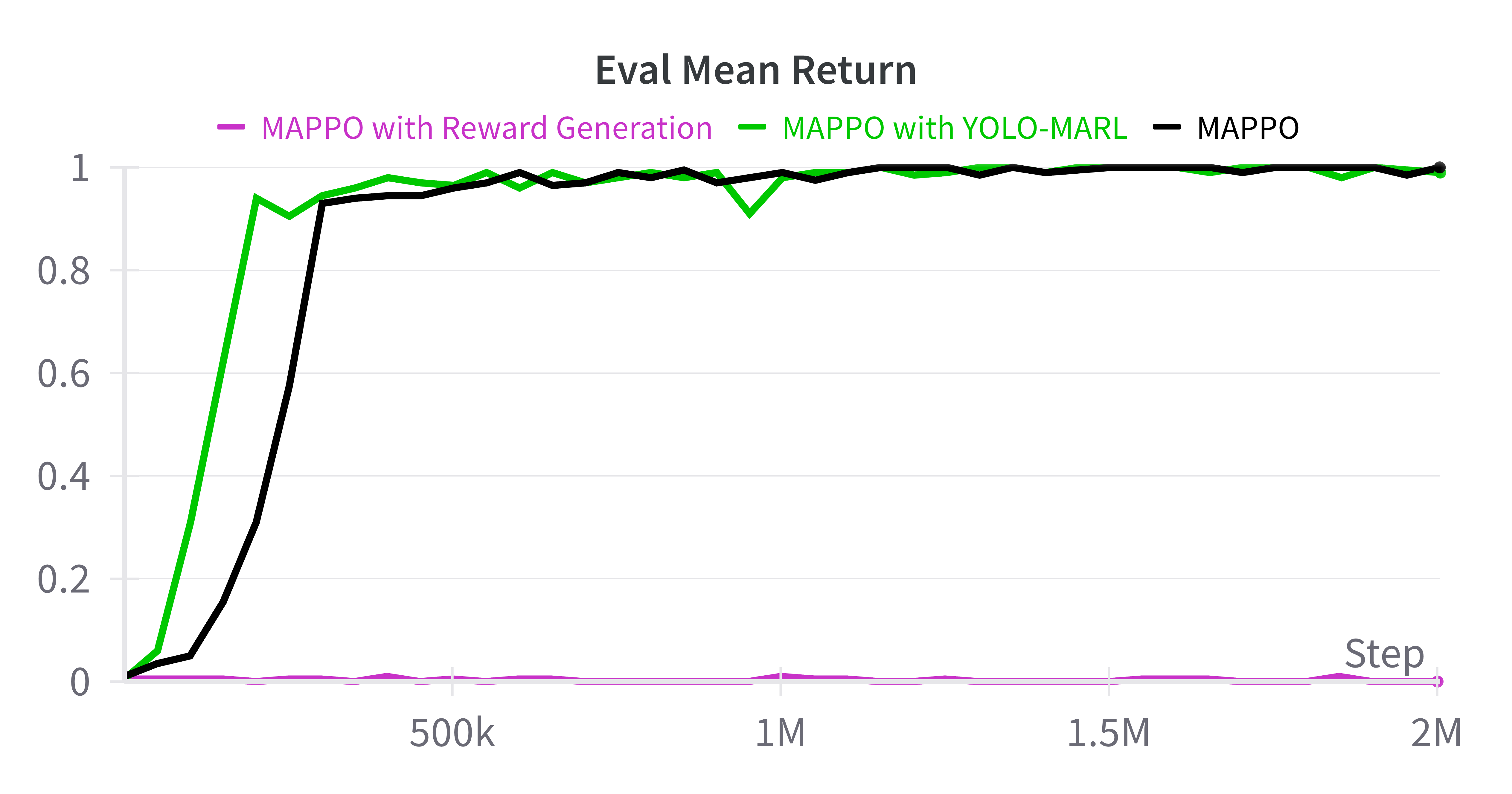}
        \caption{MAPPO}
        \label{fig:MAPPO_only_reward}
    \end{subfigure}
    \hfill
    \begin{subfigure}[b]{0.32\linewidth}
        \centering
        \includegraphics[width=\linewidth]{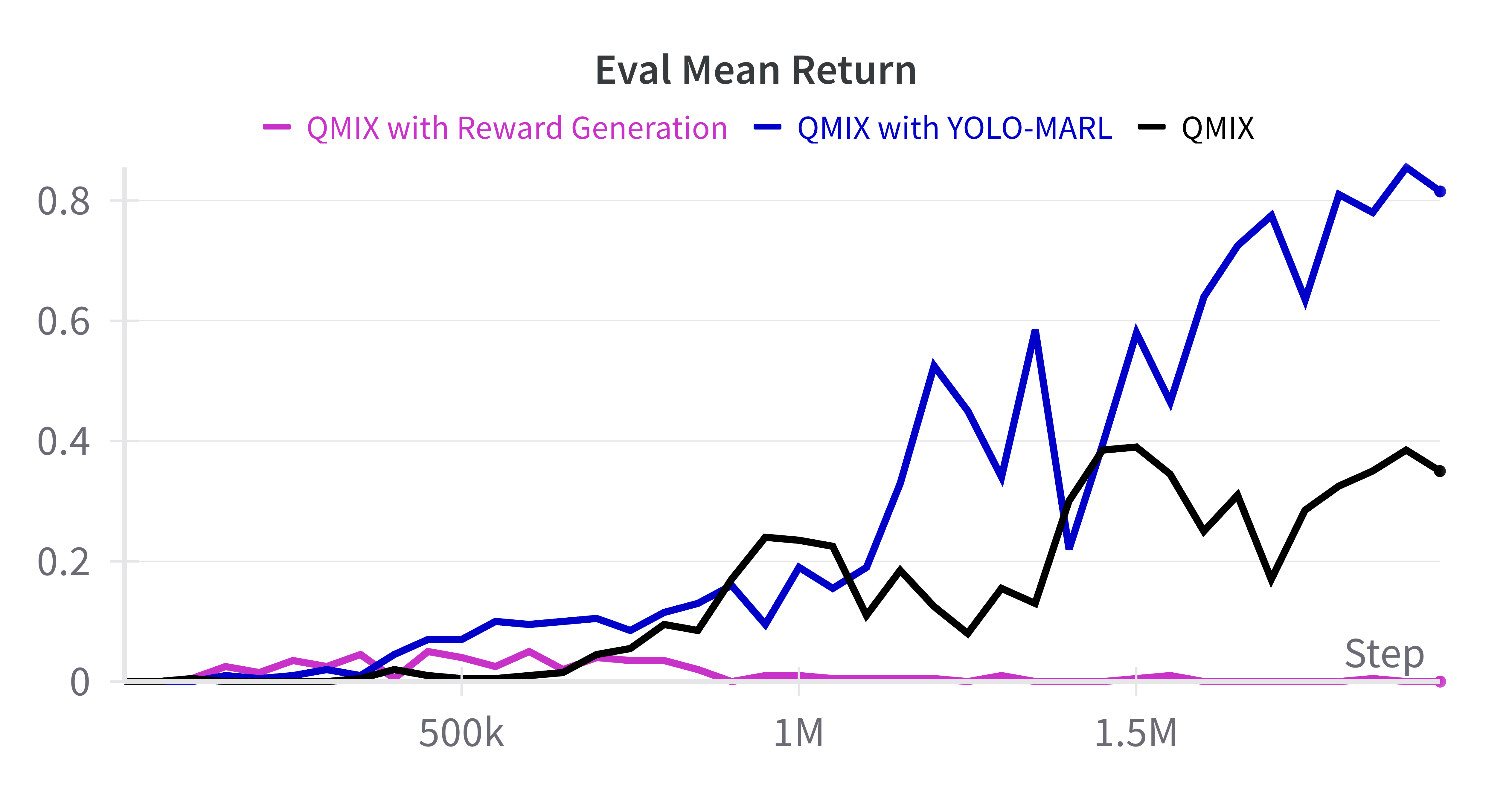}
        \caption{QMIX}
        \label{fig:QMIX_only_reward}
    \end{subfigure}
    \caption{Comparison between YOLO-MARL and reward generation without feedback in LBF}
    \label{fig:comparison_LBF_reward_only}
\end{figure*}

Our experiments show that relying solely on the LLM-generated reward function leads to poor performance. As shown in Figure~\ref{fig:comparison_LBF_reward_only}, the mean return for the LLM-generated reward function pair consistently falls below the performance of all three MARL algorithms. This indicates that agents are not learning effectively under the LLM-generated reward function. However, we do observe a slight positive return. This suggest the potential of using this framework for reward shaping tasks, particularly in situations where standard MARL algorithms struggle to learn in sparse reward scenarios. 
To investigate whether iterative refinement could improve the LLM generated reward function, we supply the LLM with the generated reward function from the prior iteration and feedback on its performance. Despite this iterative process, the LLM still fails to output a suitable reward function for the LBF environment. The mean return of evaluations remains close to zero, as shown in figure~\ref{fig:comparison_LBF_reward_feedback}.
\section{Conclusion}
We propose YOLO-MARL, a novel framework that enhances MARL policy training by integrating LLMs’ high-level planning capabilities with a single interaction per environment. This design reduces computational overhead and mitigates instability issues associated with frequent LLM interactions during training. This approach not only outperforms traditional MARL algorithms but also operates independently of the LLM during execution, demonstrating strong generalization capabilities across various environments.

We evaluate YOLO-MARL across two different environments: the MPE environment and the LBF environment. Our experiments showed that YOLO-MARL outperforms or achieve competitive results compared to baseline MARL methods. The integration of LLM-generated high-level assignment planning functions facilitated improved policy learning in challenging cooperative tasks. Finally, we mention a possible way to incorporate reward generation to our framework and we will step further.
\bibliographystyle{IEEEtran}
\bibliography{iros2025}
\addtolength{\textheight}{-12cm}   % This command serves to balance the column lengths

% \newpage
% \bibliography{iros2025}
\section{Appendix}
% \subsection{Limitation and Future Work}
% \label{sec:future_work}
% We acknowledge that the performance of YOLO-MARL may be highly correlated with the LLM's ability since even with the exact same prompt the different LLMs will output different answer. For future work, we are enthusiastic about the potential for LLMs to further enhance MARL, particularly as their planning capabilities improve. Specifically, we envision combining reward generation with planning functions to boost the performance of existing MARL algorithms in fully sparse environments.
% \subsection{Additional Results}
Given the page constraints, we present some additional experiments in this section.
\clearpage
% \begin{table*}[ht]
% \caption{Comparison between YOLO-MARL and MARL in the LBF environment across three different generated planning functions. The highest evaluation return means during training are highlighted in bold. The corresponding results can be found in figure~\ref{fig:comparison_LBF_3F}. The M means one million training steps. We use two different machines to generate planning functions and run MARL and YOLO-MARL on the same machines where the planning functions are generated.}
% \label{LBF_3F_table}
% \begin{center}
% \begin{tabular}{cccc}
%  & \multicolumn{3}{c}{Mean Return after 0.2M / 0.4M / 1.5M / 2M Steps}\\
% \cmidrule(lr){2-4}  % Proper ranges for \cmidrule
%  & QMIX   & MADDPG & MAPPO \\
% \midrule
% MARL &  0.00/ 0.01/ 0.25/ 0.36 & 0.08/ 0.28/ 0.24/ 0.29 & 0.38/  0.74/ 0.99/ 0.99\\
% YOLO-MARL & \textbf{0.00/ 0.03/ 0.69/ 0.95} & \textbf{0.18/ 0.40/ 0.42/ 0.47} & \textbf{0.94/ 0.97/ 0.99/ 0.99}\\
% \end{tabular}
% \end{center}
% \end{table*}
% \label{sec:3F}
% \vspace{2cm}
\begin{figure*}[hb]
    \centering
    \captionof{table}{Comparison between YOLO-MARL and MARL in the LBF environment across three different generated planning functions. The highest evaluation return means during training are highlighted in bold. The corresponding results can be found in figure~\ref{fig:comparison_LBF_3F}. The M means one million training steps. We use two different machines to generate planning functions and run MARL and YOLO-MARL on the same machines where the planning functions are generated.}\label{LBF_3F_table}
\begin{tabular}{cccc}
 & \multicolumn{3}{c}{Mean Return after 0.2M / 0.4M / 1.5M / 2M Steps}\\
\cmidrule(lr){2-4}  % Proper ranges for \cmidrule
 & QMIX   & MADDPG & MAPPO \\
\midrule
MARL &  0.00/ 0.01/ 0.25/ 0.36 & 0.08/ 0.28/ 0.24/ 0.29 & 0.38/  0.74/ 0.99/ 0.99\\
YOLO-MARL & \textbf{0.00/ 0.03/ 0.69/ 0.95} & \textbf{0.18/ 0.40/ 0.42/ 0.47} & \textbf{0.94/ 0.97/ 0.99/ 0.99}\\
\end{tabular}

\vspace{1cm}

    \begin{subfigure}[b]{0.32\linewidth}
        \centering
        \includegraphics[width=\linewidth]{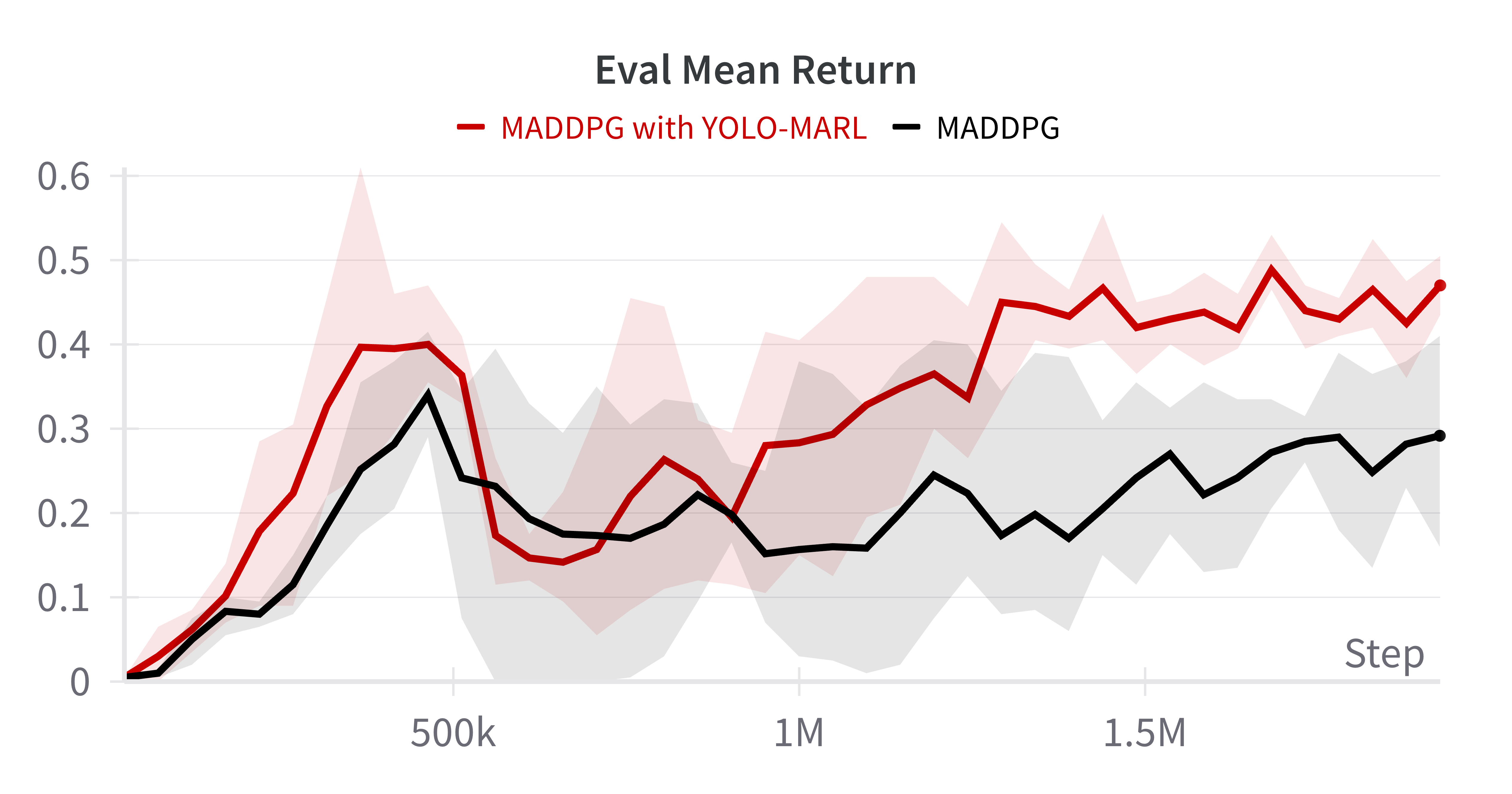}
        \caption{MADDPG}
        \label{fig:maddpg_lbf_3F}
    \end{subfigure}
    \hfill
    \begin{subfigure}[b]{0.32\linewidth}
        \centering
        \includegraphics[width=\linewidth]{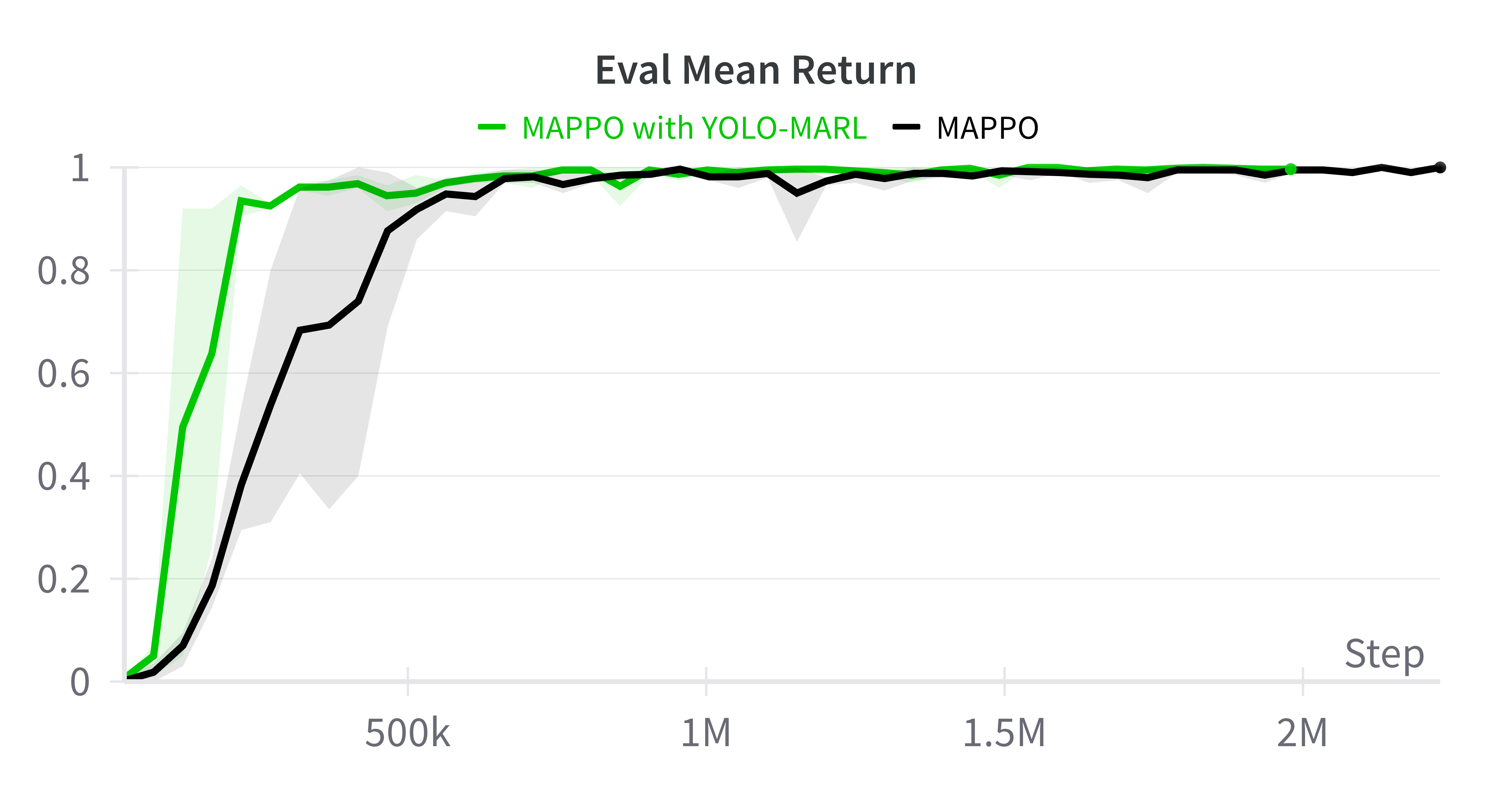}
        \caption{MAPPO}
        \label{fig:mappo_lbf_3F}
    \end{subfigure}
    \hfill
    \begin{subfigure}[b]{0.32\linewidth}
        \centering
        \includegraphics[width=\linewidth]{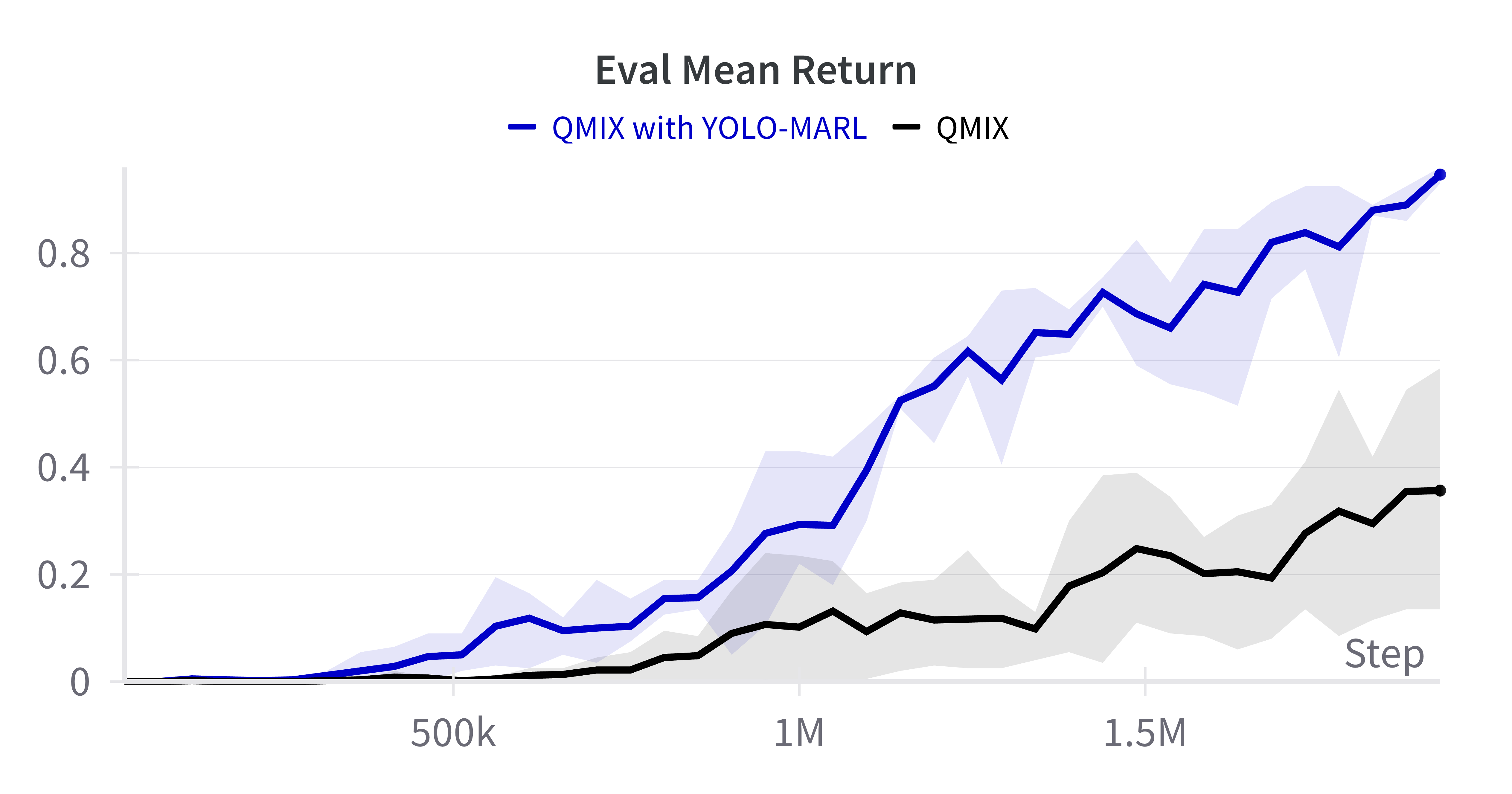}
        \caption{QMIX}
        \label{fig:qmix_lbf_3F}
    \end{subfigure}
    \captionof{figure}{\textbf{Results for LBF environment across 3 seeds:}  The solid lines indicate the mean performance, and the shaded areas represent the range (minimum to maximum) across 3 different seeds.}
    \label{fig:comparison_LBF_3F}
    
    \bigskip
    \begin{subfigure}[b]{0.4\linewidth}
        \centering
        \includegraphics[width=\linewidth]{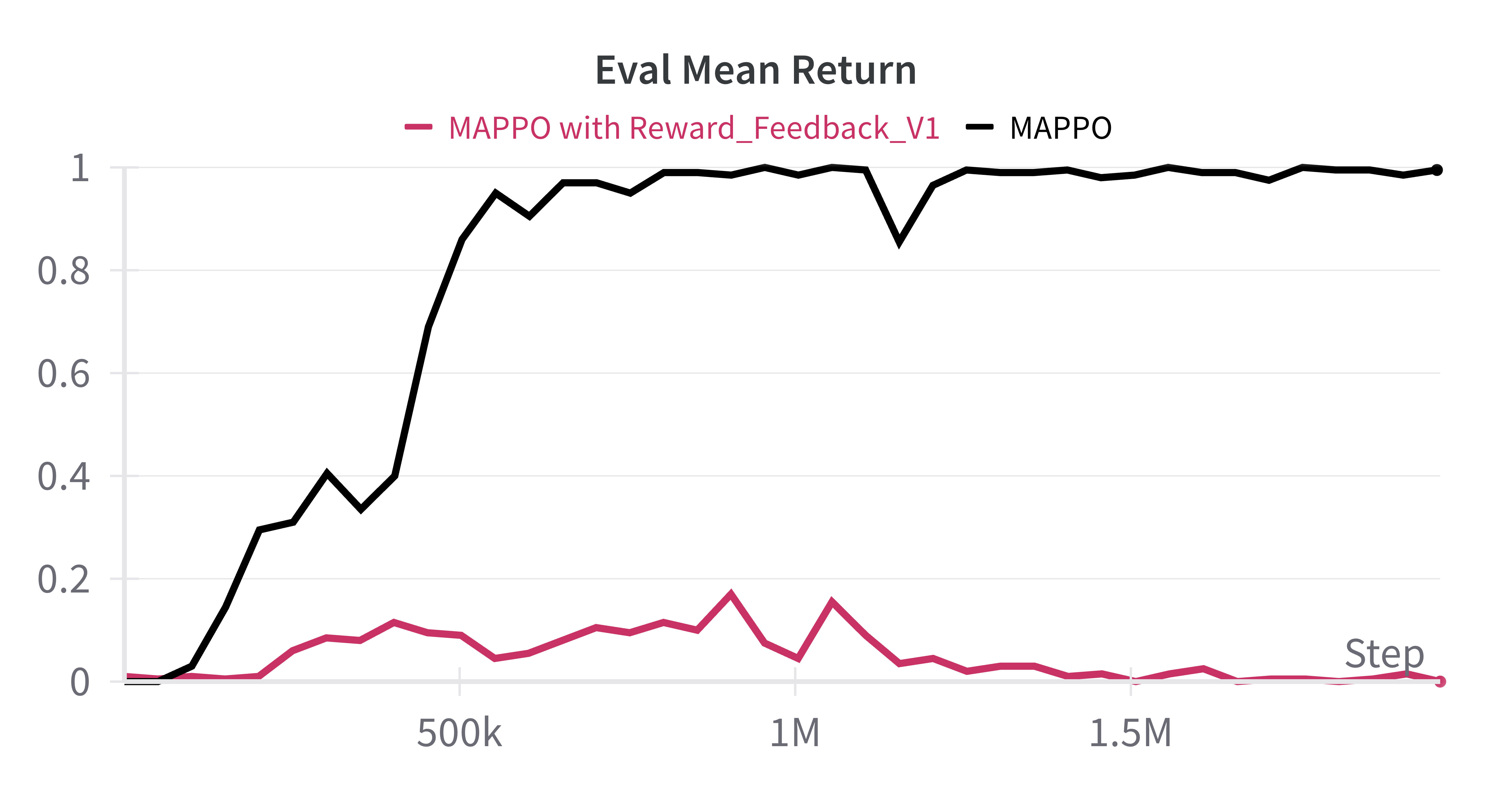}
        \caption{Iteration one}
        \label{fig:LBF_feedback_V1}
    \end{subfigure}
    \hfill
    \begin{subfigure}[b]{0.4\linewidth}
        \centering
        \includegraphics[width=\linewidth]{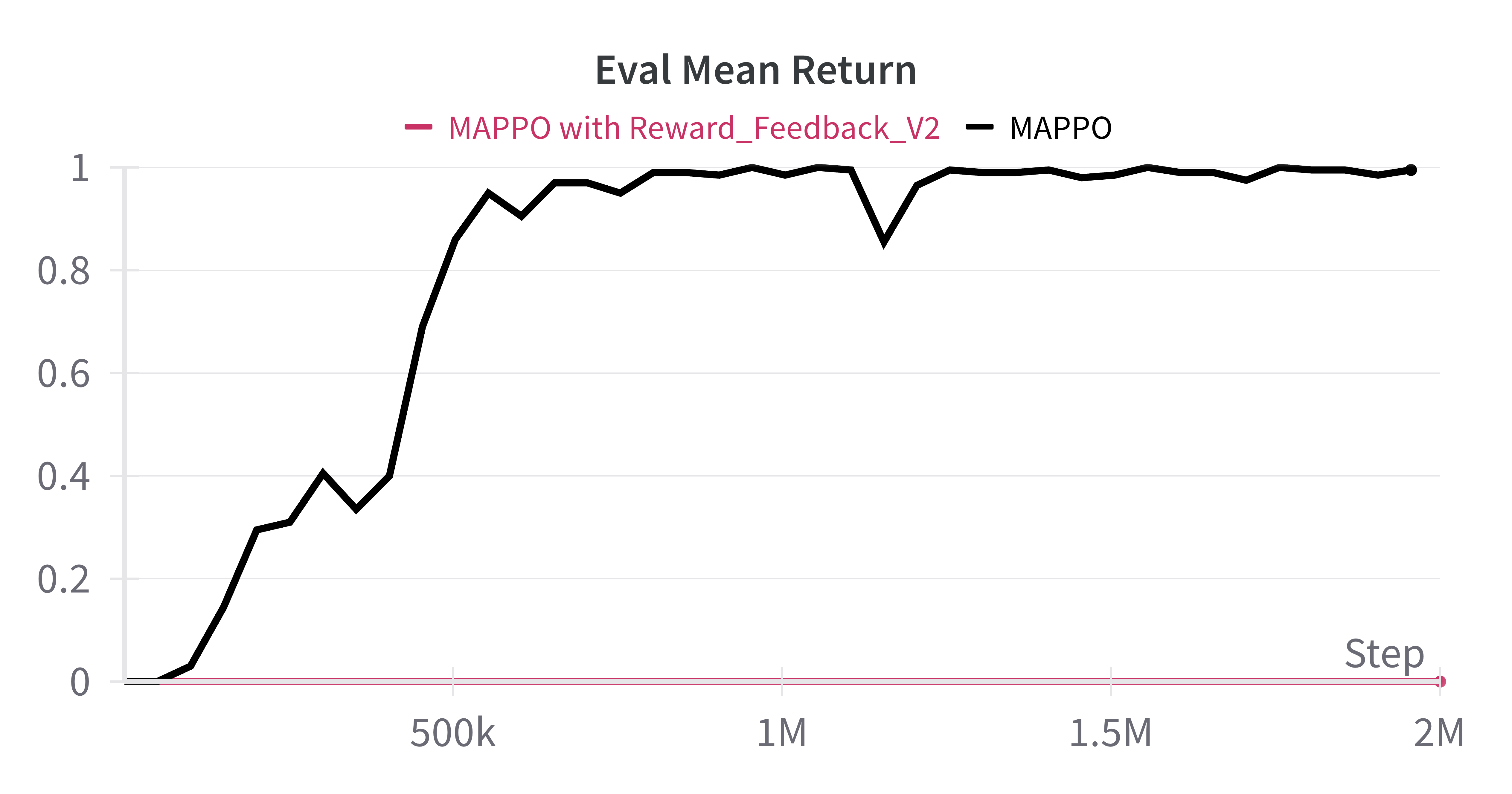}
        \caption{Iteration two}
        \label{fig:LBF_feedback_V2}
    \end{subfigure}
    \hfill
    \begin{subfigure}[b]{0.4\linewidth}
        \centering
        \includegraphics[width=\linewidth]{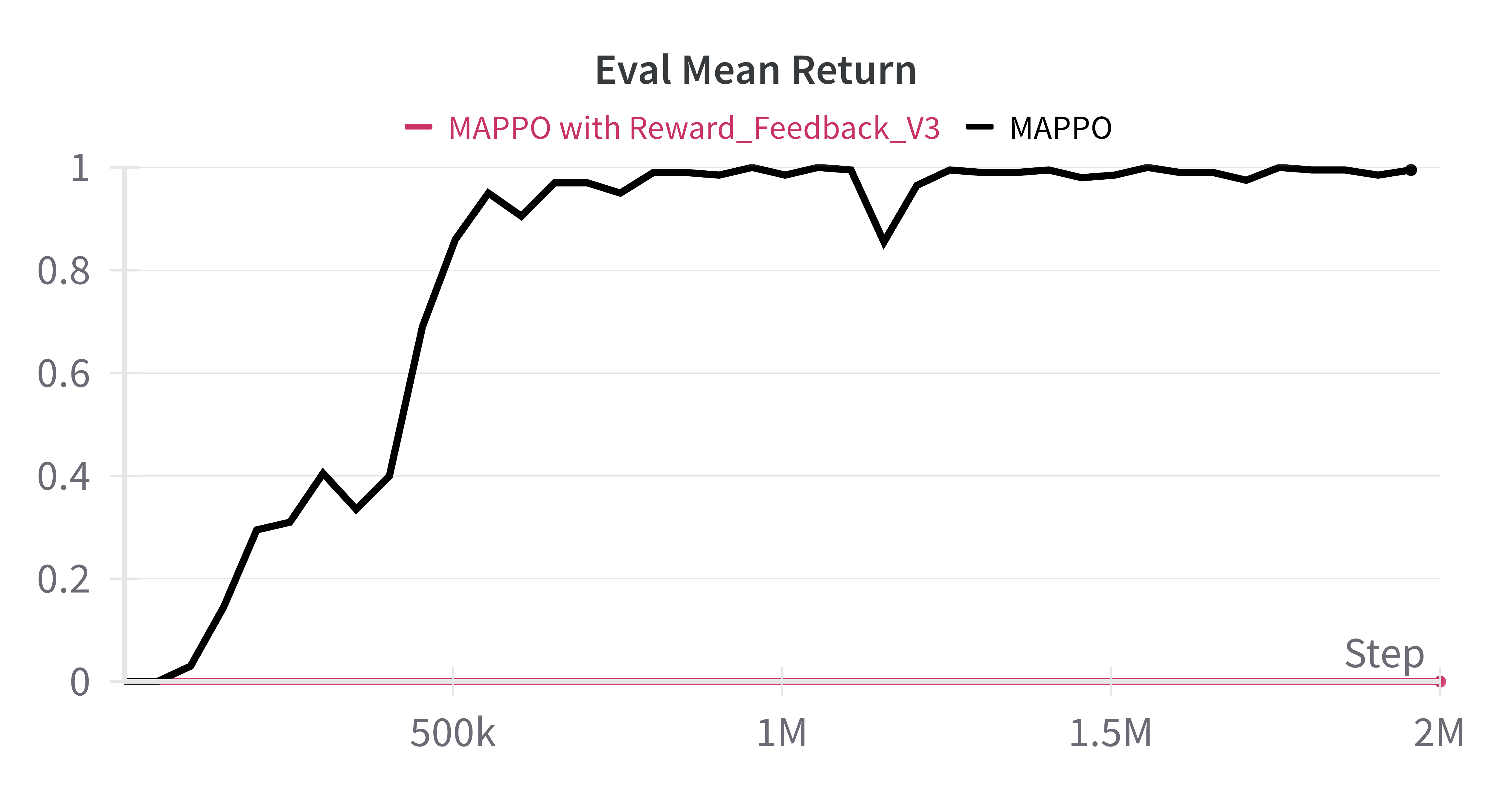}
        \caption{Iteration three}
        \label{fig:LBF_feedback_V3}
    \end{subfigure}
    \hfill
    \begin{subfigure}[b]{0.4\linewidth}
        \centering
        \includegraphics[width=\linewidth]{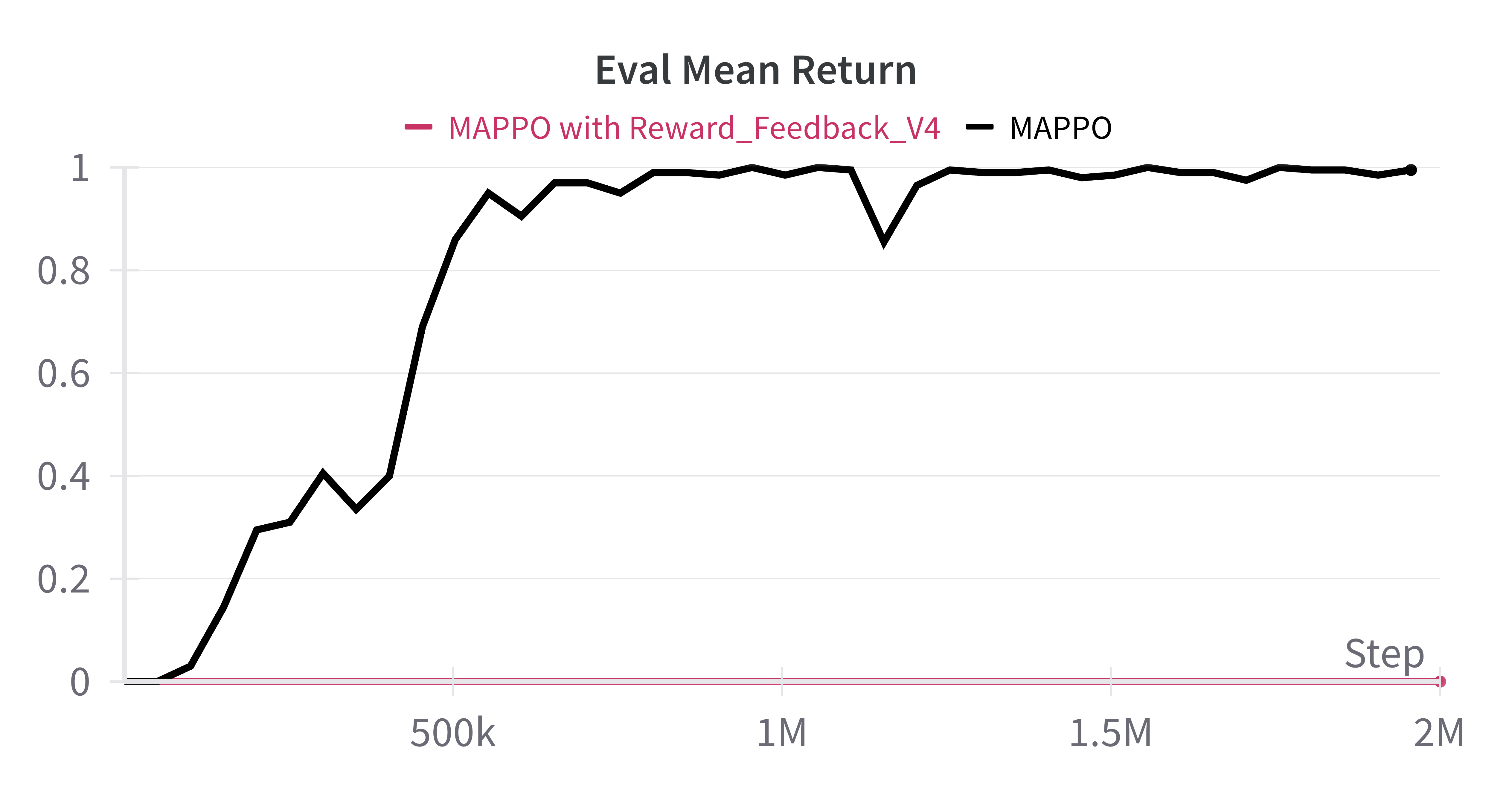}
        \caption{Iteration four}
        \label{fig:LBF_feedback_V4}
    \end{subfigure}
    \captionof{figure}{Results of only reward generation with feedback in the LBF environment. The total number of iteration is 4 and the MARL algorithm we used here is MAPPO.}
    \label{fig:comparison_LBF_reward_feedback}
    
    % \bigskip
    % \begin{subfigure}[b]{0.8\linewidth}
    %     \includegraphics[width=\linewidth]{figure/without_interp.png}
    %     \caption{Failure Case: Without providing interpretation code}
    %     \label{fig:maddpg_lbf_no_strate_interpretation}
    % \end{subfigure}
    % \hfill
    % \begin{subfigure}[b]{0.8\linewidth}
    %     \includegraphics[width=\linewidth]{figure/feedcode.png}
    %     \caption{Failure Case: Feeding environment code directly}
    %     \label{fig:feedcode}
    % \end{subfigure}
    % \hfill
    % \captionof{figure}{Failure cases for YOLO-MARL without State Interpretation Module}
    % \label{fig:comparison_no_interp}
    
\end{figure*}

\end{document}